\documentclass[aps,preprint,aps,showpacs]{revtex4}
\usepackage{subfigure}
\usepackage{epsfig}
\usepackage{graphicx}
\usepackage{amsmath}
\usepackage{amsfonts}
\usepackage{amssymb}%
\begin{document}
\title{Hybrid functional study of proper and improper multiferroics}

\author{A. Stroppa}
\affiliation{Consiglio Nazionale delle Ricerche\\
Istituto Nazionale di Fisica della Materia (CNR-INFM), CASTI Regional Laboratory,\\
67010 L' Aquila, Italy}
\author{S. Picozzi}
\affiliation{Consiglio Nazionale delle Ricerche\\
Istituto Nazionale di Fisica della Materia (CNR-INFM), CASTI Regional Laboratory,\\
67010 L' Aquila, Italy}

\begin{abstract}
 We present a detailed study of the structural,
electronic, magnetic and
ferroelectric properties of prototypical \textit{proper}
and \textit{improper} multiferroic (MF) systems such as
 BiFeO$_{3}$ and orthorhombic HoMnO$_{3}$, respectively, within density functional theory (DFT) and using the Heyd-Scuseria-Ernzerhof hybrid functional (HSE). By comparing our results with available experimental data as well as with state-of-the-art GW calculations,
 we show  that the HSE formalism is able to account well for the relevant properties of these compounds and it emerges as an accurate tool for
predictive first-principles investigations
on multiferroic systems. We show that  effects beyond local and semilocal DFT approaches
(as provided by HSE) are necessary for a realistic description of MFs.
For the electric polarization, 
a decrease is found for
  MFs with magnetically-induced ferroelectricity, such as HoMnO$_3$,  where the calculated polarization changes from $\sim$ 6 $\mu  C/cm^2$ using Perdew-Burke-Ernzerhof (PBE)  to $\sim$ 2 $\mu  C/cm^2$ using HSE.
However, for \textit{proper} MFs, such as BiFeO$_{3}$, the polarization slightly increases upon introduction of exact exchange. Our findings therefore suggest that a general trend for the HSE correction to bare density functional
cannot be extracted; rather, a specific investigation has to be carried out on each compound.
\end{abstract}

\maketitle

\section{Introduction}
Multiferroics (MFs) are materials in
which different ferroic orders  such as ferromagnetism,
ferroelectricity and/or  ferroelasticity may coexist in a single
compound.\cite{Khomskii} They have attracted much attention for
their potential applications in memory devices and other electronic
components, due to the intriguing
 possibility of controlling magnetism by an applied
  electric field, and
viceversa (magnetoelectric effect).\cite{MF1,MF2,MF3,bfo}

Multiferroics are compounds where electron correlations are rather
important, and where the electron charge shows atomic-like features,
such as strong space localization, poorly dispersed band energies,
and large on-site Coulomb repulsion.\cite{MF4,MF2,MF3,MF7} For these
systems, there are well-known deficiencies of local-spin-density
approximation (LSDA) or spin-polarized
generalized-gradient-approximation (SGGA) to
density-functional-theory (DFT). Among them, we recall the
  underestimation of the band-gap magnitude for most insulating
 materials.\cite{failures,MetallicGe,hexYMnO3}
Part of these failures can be traced back
to  the self-interaction error in  approximate
density functionals: the electron charge experiences a spurious interaction
with the Coulomb and exchange-correlation potential generated by itself.\cite{ReviewGW1,ReviewGW2}

The LSDA+$U$\cite{ldau1,ldau2,ldau3} and the self-interaction  correction (SIC) schemes\cite{sic1,sic2,sic3}  can overcome some of the deficiencies of LSDA. However, LSDA+$U$ suffers
ambiguities in the choice of the $U$ parameter 
and needs a choice regarding which orbitals to treat within a Hubbard-like approach. For simple materials, a self-consistent evaluation of the $U$ parameter can be obtained, although this method is not widely used.\cite{ldau.kresse} For BiFeO$_{3}$, the value U$_{eff}$=3.8 eV has been recently calculated.\cite{Ueff}

SIC-schemes are not commonly available in electronic structure codes
for extended solid state systems.  The implementation of a fully self-consistent
SIC-LSDA approach for extended systems was
done by Svane and co-workers.\cite{Svane} Since then,
other approaches have been implemented (for a review, see Refs.\cite{sic2,AtomicSic}). SIC-schemes  suffer from the ``nonvariationality-problem'' of the energy functional which makes forces and stress calculation not commonly available.\cite{Stengel} 

In the last few years, hybrid Hartree-Fock density functionals\cite{hybrid2,hybrid3,hybrid4,hybrid5} have been
widely used in solid state physics,\cite{hybrid5,hybrid6,hybrid7,hybrid7.5} ranging from simple semiconductor systems,\cite{hybrid7.5} to transition metals, lanthanides, actinides,\cite{itinerant1,itinerant2,itinerant3,itinerant4} molecules at surfaces,\cite{Stroppa1,Stroppa2} diluted magnetic semiconductors,\cite{Stroppa3} carbon nanostructures.\cite{hybrid11,hybrid12} For  a recent review see
Ref.\cite{ScuseriaReview}.  Hybrid functionals 
mix  the exact nonlocal exchange
of Hartree-Fock theory\cite{hybrid2,hybrid3,hybrid4,hybrid5} with the density functional exchange.
The Heyd-Scuseria-Ernzerhof screened hybrid functional (HSE)\cite{hsedef,hsedef1}
 is well suited for extended solid state systems. 

There are very few studies  dealing with ferroelectric oxides  and even less
with multiferroics. Wahl \textit{ et al.} re-investigated the well-known SrTiO$_{3}$ and BaTiO$_{3}$ \cite{Roman}
 using   HSE and semilocal functionals (LDA,PBE,\cite{pbe} PBEsol\footnote{PBEsol\cite{PBEsol} is a variant of the PBE exchange functional which improves equilibrium properties of densely-packed solids and their surfaces.}).
Bilc \textit{et al.}
studied in great details BaTiO$_{3}$ as well as  PbTiO$_{3}$ using the B1-WC
 hybrid-functional and concluded that the latter gives
 an accurate description of both the structural
and electronic properties.\cite{Bilc}
Goffinet \textit{et al.} extended the analysis to the prototypical multiferroic bismuth ferrite   showing that hybrid-functionals, specifically the B1-WC functional, open
new perspectives for a better first-principles description
of multiferroics.\cite{Goffinet} In passing, we recall that the WC gradient corrected functional\cite{WC} is very similar to PBEsol and the hybrid B1-WC\cite{B1WC}  functional mixes the WC
functional with 16\% nonlocal exchange. The hybrid HSE functional mixes 25\% nonlocal exchange with the PBE functional and the mix is performed only on the short range component of the Coulomb interaction
(for further details, we refer to Ref.\ \onlinecite{Roman}). However, which functional to prefer for simple ferroelectric compounds is still an open issue.\cite{Roman}

So far, a good performance of HSE or B1-WC functionals has
been recognized
for \textit{proper}   multiferroics where the ferroelectric polarization
 is of displacive type.
 On the other hand, magnetically
driven multiferroics, also known as \textit{improper} multiferroics,
are largely unexplored  using hybrid functionals. The purpose of this work is to extend the previous hybrid
density functional studies from prototypical ferroelectric oxides
(SrTiO$_{3}$, BaTiO$_{3}$, PbTiO$_{3}$)\cite{Roman,Bilc} or simple multiferroic system (BiFeO$_{3}$)
to more complicated and exotic multiferroic compounds, such as HoMnO$_{3}$.

 First of all, we focus on  BiFeO$_{3}$,
already investigated using the B1-WC functional, but not yet  using HSE.
In this way, we are able to compare two different, although similar, approaches for BiFeO$_{3}$. Most importantly,
we consider another prototypical case
 of \textit{improper} multiferroic,
namely HoMnO$_{3}$, which has   recently attracted
much attention.\cite{HMO1,HMO2}  We will show that important differences compared to standard DFT approaches
arise when a proper description of correlated electrons, such as that given by HSE, is taken into account.
 
Our study suggests that HSE functional improves the description compared to standard DFT approaches for multiferroic
systems.

The  paper is organized as follows. Details of the computational setups
are given in   Sect.\ \ref{Computational}. 
An extended discussion of the
structural, electronic, magnetic and ferroelectric properties of BiFeO$_{3}$ is reported in Sect.\ \ref{BFO}. Sect.\ \ref{HMOSect}
is devoted to HoMnO$_{3}$ focussing on the paraelectric AFM-A (Sect.\ \ref{A-HMO}) and ferroelectric AFM-E (Sect.\ \ref{E-HMO}) phases.
Finally, in Sect.\ \ref{conclusions}, we draw our conclusions.

\section{Computational details}\label{Computational}
All the calculations presented in this study are performed
by using the latest version of the \textit{Vienna ab initio simulation package}
(VASP 5.2).\cite{vasp1}
 For BiFeO$_{3}$, all the results are obtained using the projector-augmented plane-wave method\cite{paw1,paw2} by explicitly treating 15 valence electrons for Bi (5$d^{10}$6$s^{2}$6$p^{3}$), 14 for Fe (3p$^{6}$3$d^{6}$4$s^{2}$), and 6 for oxygen (2$s^{2}$2$p^{4}$). We used a 6$\times$6$\times$6 Monkhorst-Pack {\bf k}-mesh for the
Brillouin-zone integration and 400-eV energy cutoff. Tests using a
8$\times$8$\times$8 mesh as well as 600 eV cutoff
did not give significant differences in the calculated properties.
 Brillouin zone integrations are performed with a Gaussian broadening of 0.1 eV during all relaxations.
The experimental unit cell for the $R3c$ (ferroelectric phase)
was used as an input in the full-optimization procedure.
 For this phase as well as for  the paraelectric one (see below), we used the rhombohedral setting. The geometries were relaxed until
all force components were less than
0.01 eV/\AA\ and the stress tensor components less than 50 meV/cell. The spin configuration was fixed in order to reproduce the G-type antiferromagnetic state of BiFeO$_{3}$ and the spin-orbit coupling was neglected. For the paraelectric phase,
we used the non-polar R$\bar{3}$c LiNbO$_{3}$ phase.\cite{paraelectric}
We compute the \textit{difference} of electric polarization, \textit{i.e.} $\Delta P=P^{FE}-P^{PE}=
(P^{FE}_{ion}+P^{FE}_{ele})-(P^{PE}_{ion}+P^{PE}_{ele})=
\Delta P_{ion}+\Delta P_{ele}$, where $FE$, $PE$, $ion$ and $ele$ denote ferroelectric, paraelectric, ionic and electronic contribution, respectively.
For the paraelectric phase, we used the same lattice constant and rhombohedral angle of the ferroelectric one.  Note that, although counterintuitively,
$P^{PE}_{ele}$ may be different from zero, as explained in Ref.\ \onlinecite{Neaton}.  $P^{FE,PE}_{ion}$  is calculated  by summing the position of each ion in the unit cell
 times the  number of its valence electrons.
The electronic contribution is obtained by using the Berry phase formalism,
within the ``modern'' theory of polarization.\cite{berry1,berry2,berry3}

Concerning the HSE calculations, due to the high computational load,\footnote{To have an idea of the increased computational cost involved
 in the HSE calculation,
we note that, by considering the same computational setup for BiFeO$_{3}$, 
each electronic minimization step takes about 50 times more CPU time than PBE or PBE+U. 
This means that if a PBE (or PBE+U) self-consistent calculation takes 10 minutes, 
the HSE will take about 9 hours.} we always used the 400 eV and 6$\times$6$\times$6 {\bf k}-point mesh. The Fock exchange was sampled using the twofold reduced \textbf{k}-point grid (using the full grid, gives however negligible changes in the computed properties).
Finally, we performed G$_{0}$W$_{0}$ calculations\cite{GW,GW1,GW2,GW3,GW4,GW5}
 on top of the HSE electronic and ionic structure, which usually represent a good starting point for a perturbative quasiparticle excitation energies.\cite{GW1} We also included vertex correction in W via an effective nonlocal exchange correlation kernel.\cite{GW4}

For orthorhombic HoMnO$_{3}$, the $Pnma$ symmetry is chosen with the $b$ basis vector as the longest one. The paraelectric phase was simulated
using 20-atoms cell in the AFM-A spin configuration showing ferromagnetic (FM) (AFM) intraplanar (interplanar) coupling; for the ferroelectric one
we used a 40-atoms cell (doubling the previous cell along the $a$ axis) in the AFM-E spin configuration (\textit{i.e.} in-plane FM zigzag chains anti-ferromagnetically coupled to the neighboring chains with the interplanar coupling also AFM).
The energy cutoff was set to 300 eV and the Brillouin zone mesh was fixed to 4$\times$2$\times$4 and 2$\times$2$\times$4 grid for the AFM-A and AFM-E phase respectively. Ho $4f$ electrons were assumed as frozen in the core.
The experimental lattice constants were used for all the calculations but the internal positions were relaxed.
For the HSE calculations, the Fock operator was evaluated
on the down-folded \textbf{k}-point mesh. In order to assess the relative stability of the two magnetic phases,
we used the same simulation cell containing 40 atoms for both phases, increasing the cutoff to 400 eV and using a 4$\times$2$\times$4 {\bf k}-point grid.

\section{BiFeO$_{3}$: results and discussions}\label{BFO}
\subsection{Structural properties}
\begin{figure}[htp]
{\centering \includegraphics[width=.9\textwidth,angle=0,clip=true]{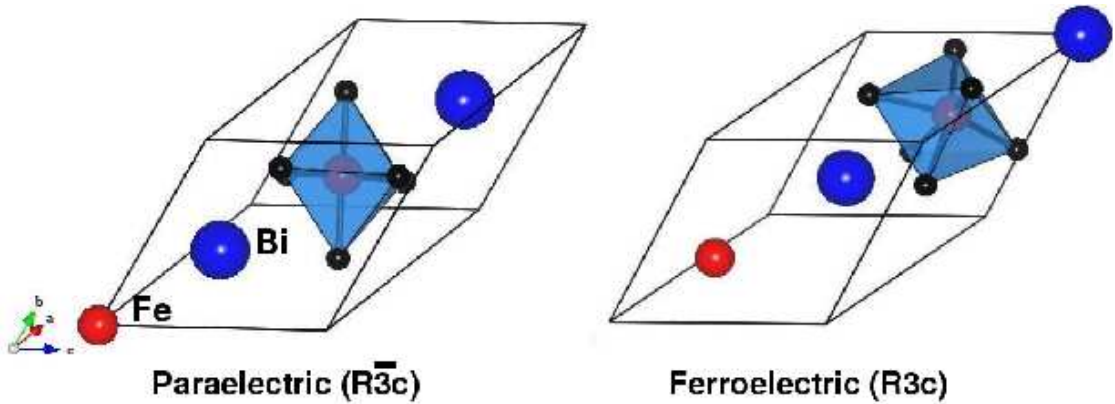}}
\caption{(Color on line) Paraelectric $R\overline{3}c$ and ferroelectric $R3c$ phase of BiFeO$_{3}$ in the rhombohedral setting. Black spheres are oxygen atoms. The octahedron centered at one Fe atom is shown. Note the off-centering of the Fe atom and the corresponding  octahedra distortion. }
\label{bfo.1}
\end{figure}

The ferroelectric structure is represented by a distorted
double perovskite structure with $R3c$ symmetry (N. 161, point group C$_{3v}$)
as reported by Kubel and Schmid.\cite{Kubel} The paraelectric phase has $R\overline{3}c$
symmetry (N. 167, point group D$_{3d}$). Both phases are shown in Fig.\ref{bfo.1}.
In Table\ \ref{tab1} we report relevant properties such as the  structural parameters, the Fe magnetic moment
 and the energy gap calculated using the PBE and HSE functionals. We also report the values using the B1-WC functional taken from Ref.\cite{Goffinet}

First of all, HSE reduces the lattice parameter $a_ {rh}$
with respect to PBE, giving a much better agreement
with the experimental value: the error decreases
from $\sim$1 \%(PBE) to $\sim$0.3 \%(HSE).
As a consequence,
 the unit cell volume $V$ also shrinks,
getting closer to the experimental value.
The rhombohedral angle, $\alpha_{rh}$,  is almost insensitive to the applied functional.  
Thus, the inclusion of  Fock exchange makes the structure more compact, \emph{i.e.}  
the lattice constant decreases.
 Note that the B1-WC functional gives too small lattice constant and too small equilibrium volume as compared to HSE, worsening the comparison with the experiments.
There is a very good agreement between the
relaxed coordinates of the
Wyckoff positions and the experimental ones
using HSE, while the PBE as well as  the B1-WC functional give slightly worse  results (
the only exception being the $x$ component of the oxygen atoms in the $6b$ site symmetry).
In the experimental structure, the BiO$_{6}$ cage is strongly distorted with three coplanar   nearest
neighbors (NNs) lying above Bi along [111] at 2.270 \AA \ (d$^{s}_{Bi-O}$, $s$ refers to short) and three
 NNs sitting below at 2.509 \AA \ (d$^{l}_{Bi-O}$, $l$ refers to long).
From Table\ \ref{tab1}, we see that the theoretical
 NNs distances  compare well with   experiments, with errors from
 $\sim$ 1 \% to  $\sim$4 \% (PBE), from $\sim$ 3 \% to $\sim$ 5 \%
(B1-WC) and  $\sim$ 2 \% (HSE).
The O-$\widehat{Fe}$-O bond angle would be
180$^{\circ}$ in the ideal
cubic perovskite. In this system, it buckles to an experimental
value of 165.03$^{\circ}$. The HSE value (164.56$^{\circ}$) is close to  PBE, and in both cases, they are slightly underestimated with respect to experiment. The B1-WC angle, on the other hand, is clearly underestimated.
Overall, the predicted HSE values clearly are in much better  agreement with experiments than those calculated using the PBE or B1-WC functional.

\subsection{Electronic and magnetic properties}
Let us consider now the magnetic and  electronic properties.
As shown  in Table\ \ref{tab1},
the calculated local moments are generally very similar for all the functionals,
and close to the experimental value. In particular,
the HSE local moment is slightly larger than   PBE , suggesting
 a more localized picture of the spin-polarized electrons.
The calculated  PBE (HSE) electronic energy gap is
 1.0 (3.4) eV. The expected band-gap opening   using hybrid
functionals can be understood as follows:
the exact exchange  acts on occupied states only, correcting them for the self-interaction, thus
shifting downwards the occupied valence bands. In turn, this has a clear interpretation: within the Hartree-Fock approximation for the ground state
of an N electron system, the potential felt by each of the N electrons in the ground state is that due to N-1 other electrons, \textit{i.e.} they feel a more attractive ionic potential. On the other hand, for unoccupied states, the potential is that due to the N occupied orbitals, so these orbitals effectively experience  a potential from one more electron, the latter screens the ionic potential which in turn becomes less attractive. Therefore, the unoccupied states are shifted upwards, opening the gap. 

As for the experimental energy gap for BiFeO$_{3}$, the situation is not clear.  There have been several measurements of the band gap using UV-visible absorption spectroscopy and 
ellipsometry on polycrystalline BFO films, epitaxial BFO films grown by pulsed-laser deposition, nanowires, nanotubes, and bulk single crystals. Reported band-gap values vary from 2.5 to 2.8 eV.\cite{gap1,gap2,gap3,gap4,Kanai,Gao}
An estimate gives $\sim$ 2.5 eV from the optical absorption
spectra by Kanai \textit{et al.}\cite{Kanai} and Gao\textit{ et al. }\cite{Gao}. From  the theoretical side, there is a spread of values: a small gap of 0.30-0.77 eV using LSDA,\cite{Neaton}
 or from 0.3 to 1.9 eV using ``LDA+U'', depending on the value of
U\cite{Neaton}; 0.8-1.0 eV using PBE (WC) GGA functional;\cite{Bilc}
3.0-3.6 eV using B1-WC and B3LYP hybrid functionals.\cite{Bilc}
Thus, a parameter-free 
 theoretical reference  value
is  clearly needed. The most accurate (but expensive) method is the GW approximation.\cite{GW} Here, we provide for the first
time, the value of the BFO energy gap based on the GW method.

First of all,  at the PBE level, we estimate an energy gap of 1.0 eV. When introducing the exact exchange (HSE),  the gap opens up to E$_{g}$=3.4  eV.
Upon inclusion of many-body effects (G$_{0}$W$_{0}$),
 it  opens even more (E$_{g}$=3.8 eV).
 Finally, when including vertex corrections, we find that   the gap reduces to 3.3 eV. Remarkably, vertex corrections almost confirm the HSE band gap.
This is perfectly in line with recent works\cite{ScuseriaReview} where it is argued  that HSE band gaps
represent a  very accurate estimate due to
partial inclusion of the  derivative discontinuity of the exchange-correlation functional.\cite{failures} Clearly, effects beyond bare DFT are important in this compound.
Our results show that hybrid-functional calculations
give already a very good estimate at a lower computational cost compared to GW. In this respect, we mention the very recent experimental study based on resonant
soft x-ray emission spectroscopy\cite{Guo-exp-gap} where  the
band gap corresponding to the energy separation between the top of the O $2p$
valence band and the bottom of the Fe $d$ conduction band is 1.3 eV.
The discrepancy between theory and experiment may be due to the presence
of defects in the experimental sample  as well as
 to the resolution involved in photoemission spectra. We hope to stimulate further experimental work to test our first-principles prediction of the
energy gap  of this important multiferroic material.

\begin{figure}
 {\centering \includegraphics[width=0.8\textwidth,angle=0,clip=true]{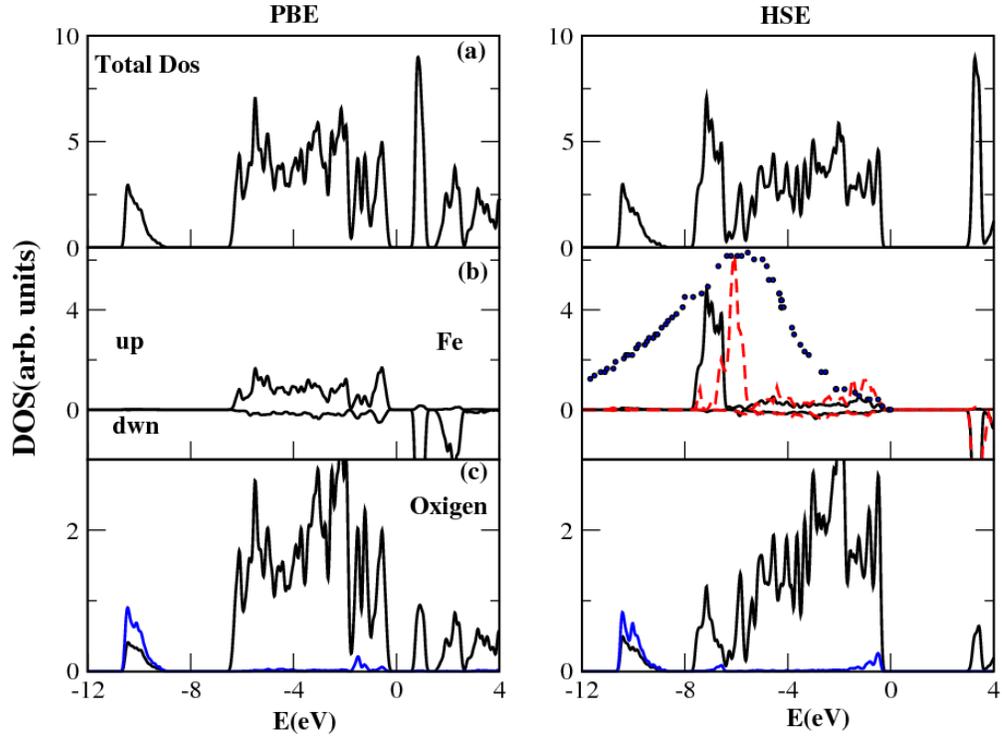}}
\caption{(Color on line) PBE (left) and HSE (right) density of states for the ferroelectric phase of BiFeO$_{3}$. In black lines: (a) total DOS for one spin channel, (b) local Fe Dos for both spin components (minority spin are shown as negative), (c) local oxygens DOS  for one spin component only. Additionally, in HSE (b), we show the Fe $d$ DOS calculated using vertex corrections in $W$ (see text) by red dashed line  and the experimental
spectral  weight of $d$ Fe taken from Ref.\cite{Guo-exp-gap} by dotted blue lines; in (c), the magenta dashed line is Bi lone-pair DOS, \textit{i.e.} Bi $s$ states.  The zero is set to valence band maximum.}\label{fig1}
\end{figure}

In Fig.\ \ref{fig1} we show the Density of States (DOS) for the optimized
atomic structure.
 Let's focus on the PBE DOS. The lowest states at $\sim$ $-$ 10 eV
are Bi $s$  hybridized with O $p$ states (blue curve).
 Above $-$ 6 eV  there are hybridized O $p$ and Fe $d$ states. The Fe $d$ states extend  in the conduction band as well; the Bi $p$ states can be found above 4 eV.  As for the HSE DOS, we see that  the conduction bands are shifted upwards, opening the valence-conduction gap.
 There is  a change in the spectral distribution  above $-$ 8 eV: a valley appears around $-$ 6 eV  and the lower shoulder of the peak increases its spectral weight. It is easy to trace back the above changes to modifications of the  majority Fe $d$ states, as shown in the panel beneath: while in PBE the band states in the vicinity of the top of the valence band have predominantly $d$ character, in HSE the spectral weight of the Fe $d$ states is concentrated far away from the top of the valence band. This can be interpreted as a change from a more itinerant picture to a more  localized description of the Fe $d$ states  going from PBE to HSE.
In Fig.\ \ref{fig1}, we also include the spectral distribution of the Fe $d$  states derived from a recent experimental work\cite{Guo-exp-gap} (see dotted lines): the position of the main HSE Fe $d$ peak almost perfectly matches the experimental PDOS, although the bandwidth of the calculated DOS is different because of energy resolution,  etc. Indeed, if we include the Fe $d$ DOS calculated using G$_{0}$W$_{0}+$ vertex corrections,
the agreement between the theoretical and experimental peak position becomes excellent. 
Note that the HSE and G$_{0}$W$_{0}+$Vertex peak position are very close to each other,
 confirming the accurate HSE description of the BiFeO$_{3}$ electronic structure. 
In passing we note that while DFT+$U$ gives a better estimate 
of the Fe $d$ peak position,\cite{Neaton}
the energy gap is still underestimated with respect our GW calculation.  
 
\begin{figure}
{\centering \includegraphics[width=0.9\textwidth,angle=0,clip=true]{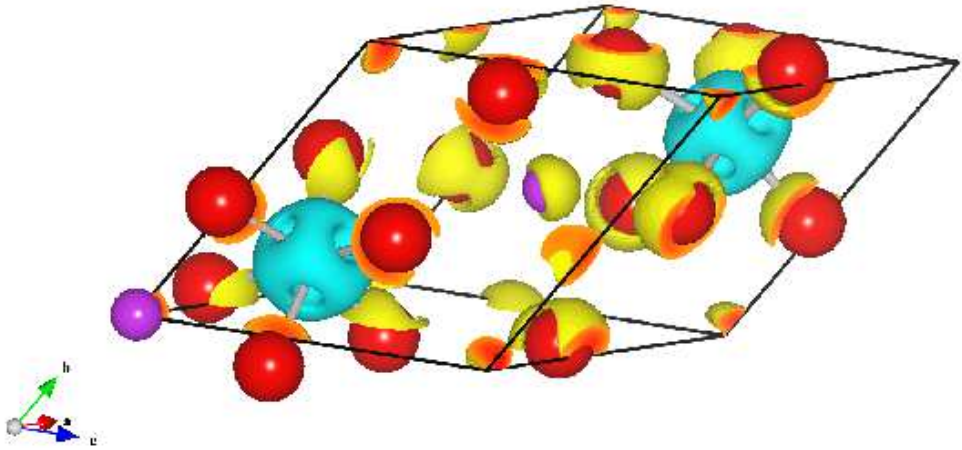}}
\caption{(Color on line) 
3D charge density difference ($\Delta \rho=\rho^{PBE}-\rho^{HSE}$) 
isosurfaces for BiFeO3.  Yellow (grey) regions correspond
 to an excess of HSE (PBE) charge. Only the distorted octahedron at Fe site is shown.}
\label{fig2}
\end{figure}

We previously mentioned that HSE may change the ionic/covalent character in this compound.
To support this, in Fig.\ \ref{fig2} we show the difference between the PBE and HSE charge density, $\Delta \rho=\rho^{PBE}-\rho^{HSE}$ calculated at fixed geometry.
In a purely ionic description, all the valence electrons would be located on the oxygens,  acting as ``electron sinks'', and the cations would donate their nominal valence charge.
The more the electrons populate the anions, or conversely, the more the electrons depopulate the cations, the more the picture shows an ionic character. Fig.\ \ref{fig2} confirms the trend discussed before: upon adding  a fraction of exact exchange to the PBE functional, the electronic charge at the cations decreases,
 \textit{i.e.} $\Delta \rho$ is positive (grey areas in Fig.\ \ref{fig2}), while the electronic charge at the anions  increases \textit{i.e.} $\Delta \rho$
is negative (yellow areas in Fig.\ \ref{fig2}). Thus, the introduction of exact-exchange generates a flux of charge from the cations towards the anions, clearly shown in Fig.\ \ref{fig2}, increasing the ionicity of the compound.

In order to discuss more quantitatively these effects,
we perform a Bader analysis
of the electronic charge.\cite{Bader1,Bader2,Bader3}
The atom in molecules (AIM) theory is a well established analysis tool for studying the topology of the electron density and suitable for
 discussing the ionic/covalent character of  a compound. The charge ($Q_{B}$) enclosed within the Bader ($V_{B}$) volume is a good approximation to the total electronic charge of an atom.  In Table\ \ref{bader} we report $Q_{B}$ and $V_{B}$ calculated for Bi, Fe, and O at a fixed geometric
structure, \textit{i.e.} HSE geometry. This is needed in order to avoid
different volumes for the normalization of the charge in the unit cell and for highlighting the electronic structure modifications due to the exact exchange. Furthermore, we consider only the valence charge
for our analysis  (although one should formally include also the core charge, we do not expect variations as far as the trends are concerned).
 Let us first consider the cations:
the Bader charge and volume are larger in PBE than in HSE. For the anions, the opposite holds true.
This is not unexpected and in agreement with intuition: upon introducing Fock exchange, the system
evolves towards a more ionic picture, through a  flux of charge from cations towards anions,
which reduces (increases) the ``size'' of the cations (anions) when going from the PBE to the HSE solution.
Finally, we note that a different degree of ionicity modifies   the calculated equilibrium lattice parameter: in a partially covalent material, such as BiFeO$_{3}$,\cite{Eriksson} the increased ionicity changes the different net charges generating a higher Madelung field,
 which is an important contribution to the bonding in the solid, and contracts the equilibrium structure.\cite{cora1,cora2}

\subsection{Ferroelectric properties}


Let's finally focus on the electric polarization. In Table\ \ref{tab4}, we report the ionic and electronic contributions to the difference of ferroelectric polarization  between the polar ($R3c$) and non-polar ($R\overline{3}c$) both in PBE and HSE. In order to disentangle the purely electronic effects from the ionic ones
upon introduction of Fock exchange, 
we report also the PBE(HSE) electronic contribution calculated at fixed HSE(PBE) geometry.

As a general comment, we note that a large polarization of $\sim$ 100 $\mu C /cm^{2}$ along (111) for bulk BFO has been reported experimentally by new measurements on high-quality single crystals.\cite{BFOexp}  in good agreement with our calculated values.
In what follows, we will mainly focus on the differences between PBE and HSE calculations. Note that the unit cell volume
  is different for PBE and HSE, as shown in Table\ \ref{tab1}.

First, we note that the polarization calculated according
to the point charge model (P$_{pcm}$) is closer to P$_{tot}$ at the
HSE than at the PBE level.
 This confirms that the HSE description of BFO points towards an  ionic picture,
 \emph{i.e.} by decreasing the covalency effects.  
The calculated total polarization P$_{tot}$ is $\sim$ 105 $\mu C/cm^2$ using PBE and $\sim$ 110 $\mu C/cm^2$ using HSE, \textit{i.e.} HSE predicts an increase of \textit{total} electric polarization. The occurrence of ferroelectricity 
 in BiFeO$_{3}$ is usually discussed in terms of ``polarizable lone pair'' carried by the Bismuth atom. This has a physical interpretation in terms of cross gap hybridization between occupied O $2p$ states and unoccupied Bi $6p$ states.\cite{Singh1,Singh2,Singh3,Singh4,Eriksson}. Intuitively, 
 the larger  the energy gap, the lower the polarization should be. Accordingly, one might expect   HSE  to reduce the polarization compared to PBE because of the larger energy gap. We will show below that this is not  in contraddiction with the results of
Table\ \ref{tab1}. In fact, let's consider P$_{tot}$ calculated at the \textit{same} atomic structure 
(for example at the PBE relaxed structure of the paraelectric and ferroelectric phases)
but using both PBE and HSE. We denote the former as  P$_{tot}^{PBE}$, the latter as P$_{tot}^{HSE(PBE)}$. Note that the ionic contributions is of course the same for both cases. From Table\ \ref{tab1}, we have P$_{tot}^{PBE}$=105.6 $\mu C/cm^2$  and  P$_{tot}^{HSE(PBE)}$=103.2  $\mu C/cm^2$, \textit{i.e.} a decrease of total polarization is found in going from PBE to HSE for the same ionic structure. A similar behavior is found in opposite conditions: for the HSE ionic structure, P$_{tot}^{HSE}$=110.3  $\mu C/cm^2$ and P$_{tot}^{PBE(HSE)}$=112.6  $\mu C/cm^2$.
Thus, keeping the same volume and including Fock exchange, the polarization reduces as   expected.
 On the other hand, when we evaluate 
the \textit{total} polarization at the appropriate equilibrium and relaxed structures using PBE and HSE , the ionic contribution also varies and one loses a direct connection between the increase of the 
energy gap  and the decrease of total polarization. In our case, the total polarization, when evaluated at the appropriate equilibrium volume for each functional, increases from PBE to HSE. This clearly points out a strong volume-dependence of the polarization, therefore calling for a correct 
estimate  of the volume (as provided by HSE).

\section{HoMnO$_{3}$:results and discussions}
\label{HMOSect}
\subsection{Paraelectric AFM-A phase} \label{A-HMO}
\subsubsection{Structural properties}
\begin{figure}
 {\centering \includegraphics[width=1.0\textwidth,angle=0,clip=true]{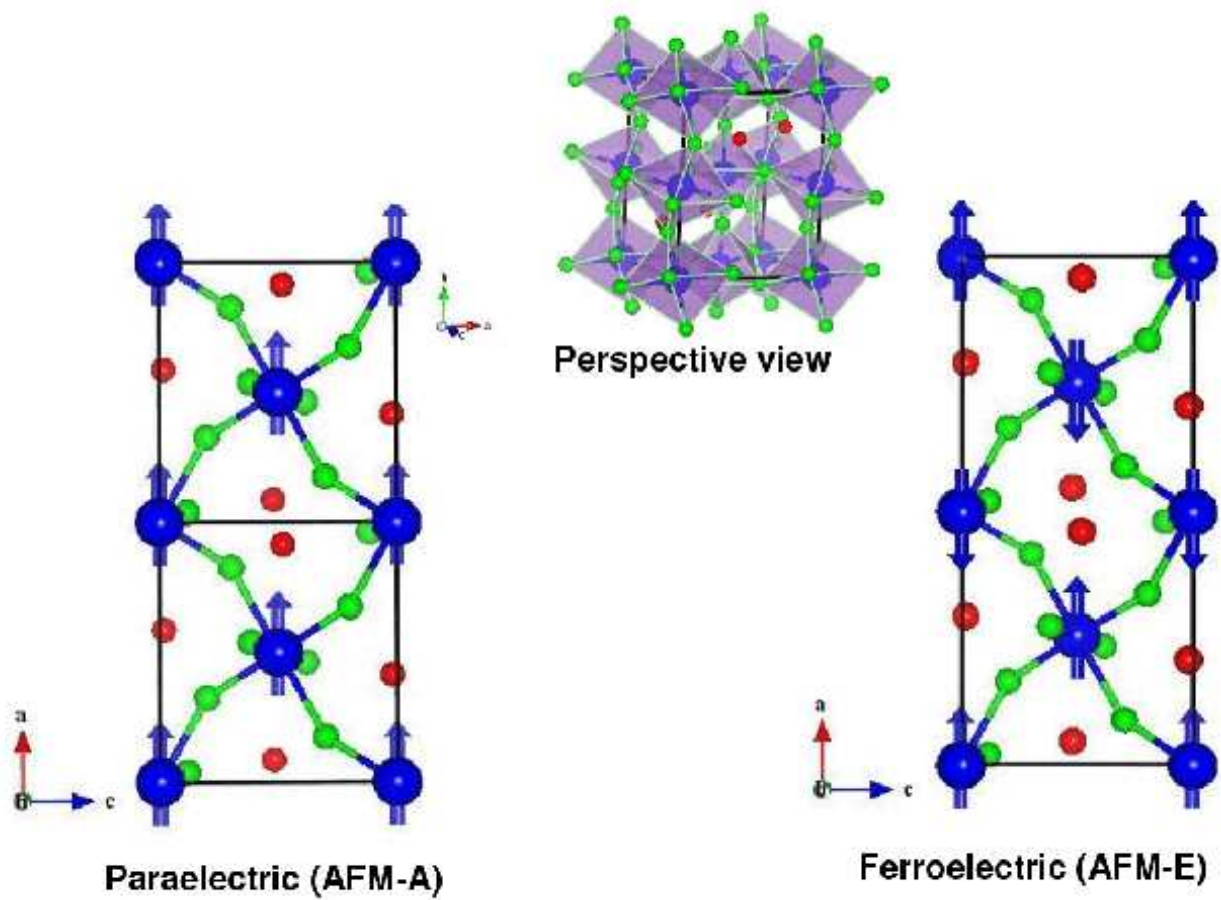}}
\caption{(Color on line) Upper part: perspective view of HoMnO$_{3}$;  polyhedra surrounding the Mn atoms are also shown.
Bottom part: paraelectric (left part) and ferroelectric (right part) spin configuration in the $c-a$ plane. The electric polarization develops along the positive $c$ axis. Red circles denote Ho atoms.}
\label{HMO}
\end{figure}

An extended review of the main properties of orthorhombic HoMnO$_{3}$
  within a standard PBE approach
can be found in Ref. \cite{HMO4} where it is also shown that the inclusion of the $U$ correction
worsens the structural properties. Therefore, in this paper,
we will focus on the comparison between the predictions of HSE with respect to  PBE results. 
In Fig.\ \ref{HMO} we show the perspective view of HoMnO$_{3}$ and the
paraelectric (AFM-A) and ferroelectric (AFM-E) spin configurations in the $c-a$ plane.

In Table\ \ref{tab.afma} we report the optimized structural
parameters in the AFM-A magnetic configuration, calculated
using  PBE and HSE.
The in-plane
short ($s$) and long ($l$) Mn-oxygen bond-lengths
get closer to experimental values using HSE; on the other hand,
the  out-of-plane length is slightly overestimated with respect to
the PBE and the experimental value. In order to quantify   structural distortions,
  the  Jahn-Teller (JT) distortion vector $\textbf{Q}=[Q_{1},Q_{2}]=[\sqrt{l-s},\sqrt{\frac{2}{3}}(2m-l-s)]$
  is often introduced.
 From Table\ \ref{tab.afma}, it is
thus clear that the HSE  functional improves the JT distortion
 upon the PBE description: the magnitude of $\textbf{Q}$ is Q=0.55 \AA \ and
0.61  \AA \ using PBE and HSE, respectively, whereas the experimental value is 0.59 \AA.
 As far as the structural angles are concerned, we first notice that the GdFeO$_{3}$-like tilting ($\alpha$) in the Mn-O$_{6}$ octahedron is slightly overestimated using the hybrid functional:
 the deviation from the experimental angle is 3.7 (PBE) and 5.0 \% (HSE) with the Mn-O-Mn in-plane  angles calculated using HSE  slightly reduced
with respect to  PBE. We note, however, that the experimental uncertainty    on the angles may be up to $\sim$ 1$^{\circ}$, \cite{HMO3}
due to synthesis problems of ortho-HoMnO$_{3}$.
\cite{HMO2,HMO3} The octahedral tilting is  related to 
 the ionic size of the rare-earth ion:\cite{Tilting,Tilt2} the tilting increases when the radius of the rare-earth atom decreases (for example, from La to Lu in the manganites series).\cite{HMO4} In this respect, the tendency towards a larger octahedral tilting, upon inclusion of exact-exchange-functional, goes hand in hand with the  reduced ionic size of Ho ion when going from PBE to HSE. As in the previous case,
we performed a Bader analysis of the valence charge distribution. Results are shown in Table\ \ref{bader}: as expected, the ''size`` of the Ho ion is reduced within  HSE. 
 According to
our previous discussion, the ionic/covalent character of the charge density is modified by HSE
in favor of a more ionic picture. This will have important consequences for the
electronic polarization, as shown below.

\subsubsection{Electronic properties}

\begin{figure}
 {\centering
 \subfigure[]
 { \includegraphics[width=0.5\textwidth,angle=-90]{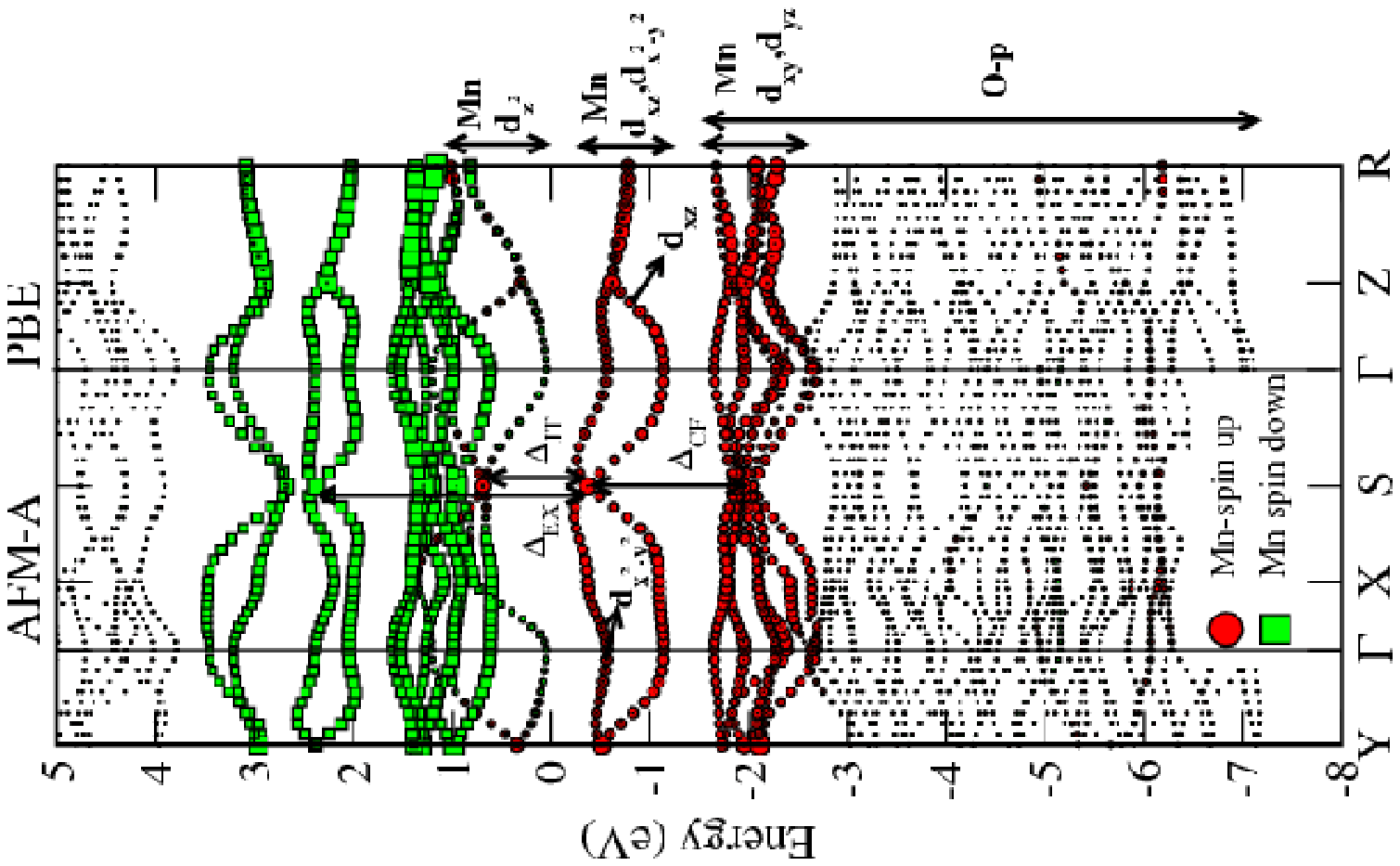}}
 \subfigure[]
 {\includegraphics[width=0.5\textwidth,angle=-90]{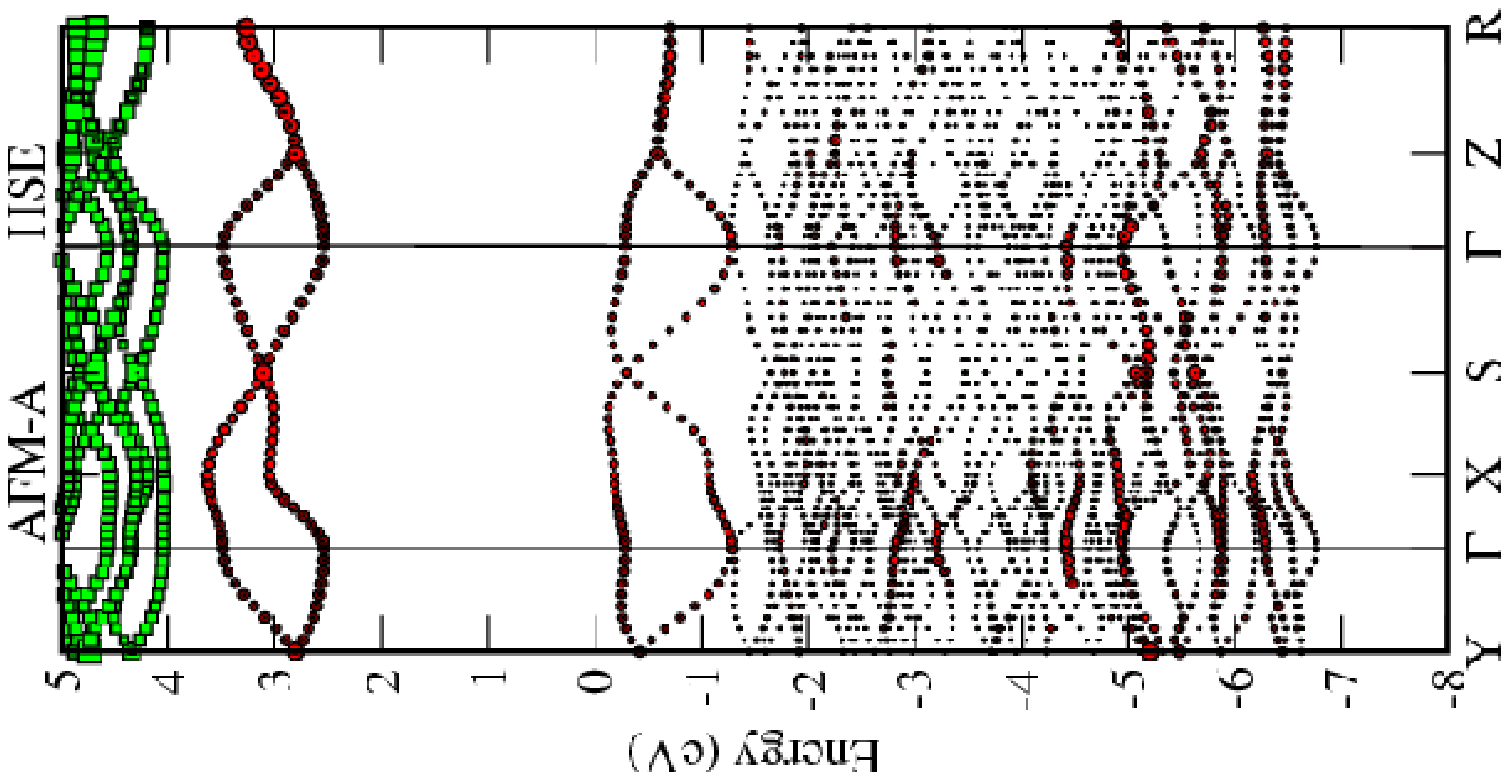}}}
\caption{(Color on line) Band structure for the AFM-A phase of HoMnO$_{3}$ calculated
along the symmetry lines of the orthorhombic Brillouin zone,
for PBE (left) and HSE (right) using the relaxed structures. Red (green) dots refer to bands projected onto  spin-up(-down) Mn atoms.}
\label{HMO.bande}
\end{figure}

In Fig.\ \ref{HMO.bande} we show the band structure for the AFM-A phase as calculated using standard PBE (left panel) as well as HSE (right panel) along the main  symmetry lines.
The PBE band structure shows a small gap equal to $\sim$ 0.2 eV. The bands below $\sim$ $-$2  eV are mainly oxygen $p$ states
and those  2-3 eV below (above) the Fermi level are mainly spin-up (spin-down)
Mn $d$ states. There is also a considerable weight of the Mn $d$ states 
in the oxygen bands near the top of the valence band.\footnote{Although the set of ``$t_{2g}$'' or ``$e_{g}$''
orbitals is well defined in a local coordinate frame centered on each Mn ion, this is not any more true when using
the standard orthorhombic system as a global coordinate frame, due to different tilting angles and distortions on neighboring MnO$_6$ cages. Thus, our discussion for the Mn $d$ states has only a qualitative meaning.}

The group of bands between $-$ 1 and $-$ 2 eV are mainly $d_{xy}$,$d_{yz}$ with some small weight of $d_{z^{2}}$.
The  two  bands just below the Fermi energy are  mainly $d_{x^{2}-y^{2}}$-like with some $d_{xz}$ weight; at   $Y$,$S$,$Z$   they become degenerate.
Higher in energy, between 0 and 1 eV,
there are two  more bands showing a similar behavior, \textit{i.e.} degenerate at $Y$,$S$,$Z$, almost degenerate along $Z$-$R$ and with a similar overall band dispersion. Even  higher in energy,
the are the  Mn minority states.
From Fig.\ \ref{HMO.bande}, we can extract the JT splitting ($\Delta_{JT}$ of the ``$e_{g}$'' states), the CF splitting ($\Delta_{CF}$ between ``$e_{g}$'' and ``$t_{2g}$''), and the exchange splitting ($\Delta_{EX}$ between majority and minority spin states, evaluated at the $S$ point, for simplicity). We thus have
 $\Delta_{JT}$=1.05, $\Delta_{CF}$=2.34, and $\Delta_{EX}$=2.75 eV.
The comparison between the PBE and HSE band structure highlights some
differences.  First we note that the oxygen bands are slightly shifted to lower binding energy together with the Mn $d_{xy}$,$d_{yz}$ bands. On the other hand, the average position of occupied Mn $d_{xz}$, $d_{x^{2}-y^{2}}$  bands remain almost
unchanged, but the band-width increases. The latter effect is mainly   shown by the lowest  of the two bands, \textit{i.e.} the $d_{xy}$-like band. As   well known,\cite{MartinBook}
Hartree-Fock hamiltonians naturally leads to larger band-width and down-shift of electronic states, even for
simple homogeneous systems. Thus, the larger band-widths obtained by HSE can not be simply  connected
to a stronger hybridization, because it is an intrinsic feature of Fock exchange for every electronic state.
Indeed, we will show below that the $d-p$ hybridization is expected to decrease using HSE.
 
$\Delta_{JT}$ is evaluated as  3.4 eV, larger than in the PBE case, suggesting a stronger local distortion related to the  Jahn-Teller instability. The increase of the JT  splitting is linked to the increase of the energy gap as well, which is now  $\sim$ 2.7 eV. The exchange splitting, $\Delta_{EX}$=4.6 eV, is also larger along with an  increased  local Mn moments with respect to PBE.
Note, that the enhancement of Jahn-Teller distortion by HSE is not unexpected. In fact,
it was previously suggested that the HSE functional is able to reveal the Jahn-Teller effect for Mn$^{+4}$ through a \textit{symmetry broken solution}
giving rise to an orbitally ordered state and consequent Jahn-Teller distortion.\cite{Stroppa3}
The larger $\Delta_{JT}$ is mainly driven by a purely electronic effect due to the inclusion of  Fock-exchange. Infact, by calculating the PBE self-consistent charge density on top of the HSE ionic
structure, $\Delta_{JT}$ becomes $\sim$ 1.2 eV, \textit{i.e.} nearly equal to the previous PBE case.
Thus, we expect  HSE to cause a rearrangement of the charge density that will \textit{reduce} the
electronic contribution to the electronic polarization when considering the polar phase.  This  can be understood as follows.
The increase of the Jahn-Teller splitting goes hand in hand  with the increase of the energy gap:  the larger  the  gap, the smaller  the  dielectric constant is, \textit{i.e.} the smaller the screening is. Now, let us consider the fixed ionic configuration of the paraelectric phase: the charge in the non-centrosymmetric \textit{spin arrangement}   can be thought as a ``small'' perturbation of the centrosymmetric one   upon the application of the  ``internal'' electric field.
The electronic charge will  respond to such a field, and  each electronic state will change assuming a polarized configuration. If the gap is large, the ``electric field'' will hardly mix the electronic states in the valence band with the electronic states in the conduction band, since in second order perturbation theory approach the denominator will be of order of the band gap energy,
 so that  the electrons don't polarize much, \textit{i.e.} the charge distribution becomes more ``rigid''. In conclusion, we expect  the electronic contribution to the electric polarization to decrease upon introduction of Fock exchange.
This will be confirmed by our calculations. Note that the above reasoning is not appropriate for BFO where the polarization is mainly due to  ionic displacements.

\subsection{AFM-E phase}\label{E-HMO}
\subsubsection{Structural properties}
\begin{figure}
{\centering
\includegraphics[width=0.9\textwidth,angle=0,clip=true]{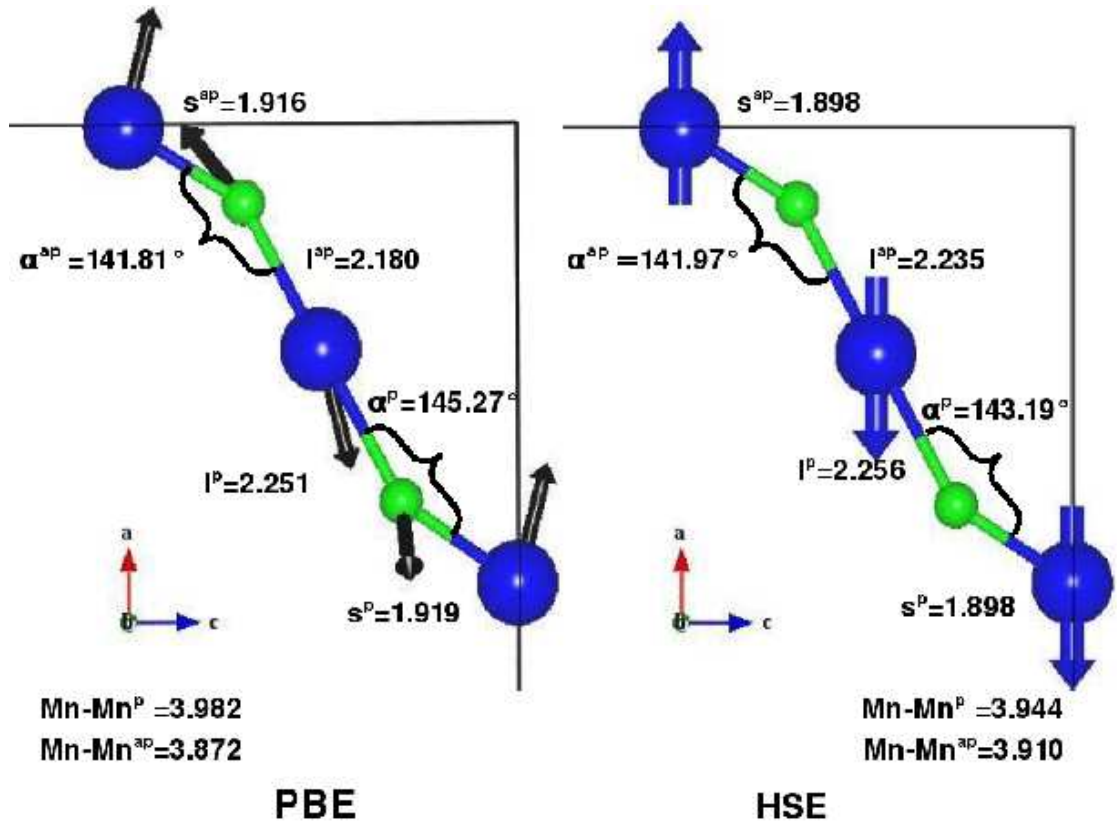}}
\caption{(Color on line) Schematic representation of the Mn-O-Mn-O-Mn chain in the $c-a$ plane, corresponding to the up-down-down spin configuration. The structures are in scale and
all structural details are shown in the Figure (see text for further details). Distances are in \AA\ and
angles are in decimal degrees ($^\circ$). The left (right) part corresponds to the PBE (HSE) relaxed structure. The atomic displacements projected into the $c-a$ plane which bring the PBE structure to the HSE one are shown by black arrows (left part). The spin configuration is shown only in the HSE structure by blue arrows. Blue (green) spheres are Mn (oxygen) atoms. Ho atoms are not shown for clarity.}
\label{chain}
\end{figure}

Let us focus on the AFM-E phase, where the resulting symmetry is lowered by
the spin configuration with respect to the AFM-A spin arrangements by removing
 the inversion symmetry.
In Fig.\ \ref{chain}, we show the relevant structural internal parameters, for the relaxed PBE and HSE structure,  by considering the Mn-O-Mn-O-Mn chain (compare Fig.\ \ref{chain} and Fig.\ \ref{HMO}).
 The Mn-O $short$  bond-lengths do not show significant differences
between the parallel and antiparallel spin configuration  in PBE as well in HSE. On the other hand, the long Mn-O bond lengths are mostly affected:
their difference, $l^{p}$ - $l^{ap}$ in PBE is about 0.07 \AA \ and decreases to $\sim$ 0.02 \AA \ upon introduction of exact exchange.
At the same time, the angle changes: $\alpha^{p}$ decreases while $\alpha^{ap}$
remains almost equal to the PBE value.
  The results can be interpreted as follows: bare PBE is expected to overestimate hybridization effects between oxygen $p$-states and Mn $d$-states, therefore inducing a stronger rearrangement of ionic positions consistent with a ``softer" structure when moving from, say a centrosymmetric A-type to a ferroelectric E-type phase. Viceversa, upon introduction of correlation effects, the reduced hybridization is expected to lead to a more ``rigid"  ionic arrangement. Indeed, this is evident when comparing
the difference between   $\alpha^{p}$ - $\alpha^{ap}$, which drastically reduces upon introduction of HSE with respect to PBE. We recall that, ultimately, it is this difference that gives rise to the ionic polarization, as clearly shown in Fig. 2 b) of Ref.\cite{HMO1}. We can therefore anticipate that a reduction of the polarization will occur upon introduction of HSE, as discussed in detail below.
What is worthwhile noting is that the Mn-Mn distances in PBE dramatically depends on their having parallel ($d_{Mn-Mn}^p$ = 3.98 \AA) or antiparallel spins ($d_{Mn-Mn}^{ap}$ = 3.87 \AA), so that
$d_{Mn-Mn}^p - d_{Mn-Mn}^{ap}$ = 0.11 \AA. However, this dependence is smoothed upon introduction of HSE, so that the difference in Mn-Mn distance strongly reduces to  $d_{Mn-Mn}^p - d_{Mn-Mn}^{ap}$ = 0.03 \AA. In general, the marked (weak) dependence of the structural properties within PBE (HSE) is consistent with a strong (small) efficiency of the double--exchange mechanism, which ultimately relies on the $p-d$ hybridization and hopping integral.

\subsubsection{Electronic and magnetic properties}
 The band-structure of the AFM-E is quite similar to the A-type and is therefore not shown. However, there are some small differences which we comment on.
 As expected, the increase of the number of the AFM bonds
of each Mn with its four neighbors  associated with the change of the magnetic state going from AFM-A to AFM-E type results
in a narrowing of all bands. This is  further enhanced by HSE due to the  reduced hopping upon introduction of exact exchange, as expected. Furthermore, the increase of the
band gap is facilitated by the interplay of the crystal distortion,
which is generally enhanced by HSE,  with
the AFM arrangement of spins. As expected, the energy gap
is the largest in the AFM-E-HSE band structure, being now $\sim$ 3 eV.
The $\Delta_{JT}$ evaluated at $S$ point, is also the largest in this case,
being $\sim$ 3.7 eV.  
Before turning our attention to the electronic polarization,
we discuss the magnetic properties. First of all,
we found that the AFM-E is more stable than the AFM-A by $\sim$ 4 meV/cell in the HSE
formalism. Note that this value has been obtained using the same simulation cell for both phases, therefore reducing the influence of numerical errors. Although the relative stability 
is still comparable with the numerical accuracy, it is indeed consistent   with experiments.\cite{HMO2,HMO3} In AFM-E-PBE, the Mn moment is 3.4 $\mu_{B}$ which induces a small spin-polarization on the oxygen equal to $\pm$ 0.04 $\mu_{B}$. In AFM-E-HSE, the Mn moment slightly increases to 3.7 $\mu_{B}$
and the oxygen moment slightly decreases to $\pm$ 0.01 $\mu_{B}$:
the increased localization of the Mn $d$ states correlates with the increased
Manganese spin moment and goes hand by hand with the  decreased $p$-$d$ hybridization and a decreased induced spin moment on oxygens.

\subsubsection{Ferroelectric properties}
Finally, we calculated the ferroelectric polarization by considering the AFM-A as the reference paraelectric structure.
The results show that the polarization (both the electronic and ionic terms) strongly reduces upon introduction of HSE. However, it is remarkable that the total P is still of the order of 2 $\mu C/cm^2$:
 this confirms HoMnO$_3$ as the magnetically-induced ferroelectric having the highest 
 polarization predicted so far. Our estimate is in very good agreement with model Hamiltonian
 calculations.\cite{dagotto}  The comparison between
theory and experiments as far as the electric polarization is concerned is still a matter of debate. Whereas earlier studies predicted negligible values for polycrystalline HoMnO$_{3}$ samples,\cite{Lorenz}  more recent studies for TmMnO$_{3}$
in the E-type (where the exchange-striction mechanism is exactly the same as in HoMnO$_{3}$) point to a polarization which could exceed 1 $\mu C/cm^2$,\cite{Yu}
in excellent agreement with our predicted HSE value.

The reasons why we expect a reduction upon introduction of HSE have been already discussed in previous paragraphs and can be traced back to the reduced $p-d$ hybridization. As in the case of BFO, we disentangle the structural and electronic effects, by using the HSE (PBE) geometry with the PBE (HSE) functional 
(cfr.  Table\ \ref{tab4}). What we infer from these ``ad-hoc"-built systems is that the use of
 HSE dramatically reduces the electronic contribution (cfr $P_{ele}$ in PBE and HSE(PBE)), {\em i.e.} reduced by $\sim$ 2.5 $\mu C/cm^2$. Less important, though still appreciable, seem to be the ionic displacements: their dipole moment is reduced by $\sim$ 1.5 $\mu C/cm^2$ when comparing $P_{tot}$ in HSE and HSE(PBE). This is consistent with what shown approximately by P$_{pcm}$.

\section{Conclusions}\label{conclusions}
In this work, we have revised the two workhorse materials of the exponentially growing
field of multiferroics, namely BiFeO$_{3}$ for proper MFs and HoMnO$_{3}$
for improper MFs by using the  screened hybrid functional (HSE).

From our study,  several important points emerge.
For BFO:
i)  the structural, electronic and magnetic properties well agree  with experiments;
ii) the ferroelectric polarization agrees with reported values in the literature;
iii) even if PBE allows the description of ferroelectric properties by
 opening an energy gap, it is by no means  satisfactory in correctly describing
 all the properties on the same footing. On the other hand, HSE improves the  PBE and LDA+$U$ description; this is clearly shown by benchmark
calculations using the most advanced and accurate
state-of-the art GW+vertex corrections (which basically confirm the HSE results);
iv) the previous comment, and very recent studies\cite{ScuseriaReview}
 suggests that optical properties, so far not investigated at all by ab--initio calculations for BFO, can be properly addressed within HSE.
v) finally, we note that the electronic polarization \textit{increases}  upon introduction of exact exchange. 
For HMO, we note that:
i) the HSE results are in good agreement with experiments when available;
ii) the  Jahn-Teller effect is correctly described in agreement with experiment;
iii) despite a reduction of the polarization value with respect to PBE, HoMnO$_{3}$ still shows the highest $P$ predicted among magnetically-induced ferroelectrics.

We have shown that
 introduction of ''correlation`` effects may both enhance the polarization
or reduce it: the former effect will most
likely occur for proper MFs, and the latter
for improper MFs, \textit{e.g.} magnetically driven.
Note that an increase of HSE polarization with respect to LDA, for example,
 is also found by Wahl \textit{et al.}\cite{Roman} for BaTiO$_{3}$, a standard ferroelectric compound. 
Also for BiFeO$_{3}$, an increase of polarization using DFT+$U$ 
has been noticed.\cite{Eriksson,Neaton,BFO.LDAU} The increase of ferroelectric polarization
when including a fraction of exact exchange and using the theoretical equilibrium 
volume has been reported also for simple ferrolectric compound such as KNbO$_{3}$.\cite{cora1}

One final comment is in order: although the HSE results certainly point
towards a truly realistic description,
it is still possible that, to some extent,
 the good performances of HSE may be material-dependent, \textit{i.e.} the universal 1/4 fraction of the exact exchange may be not appropriate for some  specific material. What is certainly true is that
the predictive capability of HSE, combined with
 its nowadays affordable computational costs, make the functional an attractive
 choice for the study of a wide range of materials,
 from well-behaved insulators to doped semiconductors
 exhibiting magnetic ordering, multifunctional complex oxides  of interest for many industrial applications therefore representing a very good starting point for materials design.

\acknowledgments
The research leading to these results has received funding from the European Research Council under the European Community,  7th Framework Programme - FP7 (2007-2013)/ERC Grant Agreement n. 203523. A.S. would like to thank  G. Kresse for kind assistance for the GW calculations and M. Marsman for useful discussions. Furthermore,
A.S. thanks  L. Kronik (Weizmann Institute) for helpful comments. 
The authors acknowledge kind hospitality at the S$^3$ CNR-INFM National Center in Modena after the catastrophic earthquake  of April 6$^{th}$ 2009 in L'Aquila.
The computational support by Caspur Supercomputing Center in Roma and technical assistance by 
Dr. L. Ferraro is gratefully acknowledged. Figures have been done by using the VESTA package.\cite{vesta}
\newpage


\begin{thebibliography}{95}
\expandafter\ifx\csname natexlab\endcsname\relax\def\natexlab#1{#1}\fi
\expandafter\ifx\csname bibnamefont\endcsname\relax
  \def\bibnamefont#1{#1}\fi
\expandafter\ifx\csname bibfnamefont\endcsname\relax
  \def\bibfnamefont#1{#1}\fi
\expandafter\ifx\csname citenamefont\endcsname\relax
  \def\citenamefont#1{#1}\fi
\expandafter\ifx\csname url\endcsname\relax
  \def\url#1{\texttt{#1}}\fi
\expandafter\ifx\csname urlprefix\endcsname\relax\def\urlprefix{URL }\fi
\providecommand{\bibinfo}[2]{#2}
\providecommand{\eprint}[2][]{\url{#2}}

\bibitem[{\citenamefont{Khomskii}(2009)}]{Khomskii}
\bibinfo{author}{\bibfnamefont{D.~I.} \bibnamefont{Khomskii}},
  \bibinfo{journal}{Physics}, \bibinfo{year}{2009}, \textbf{\bibinfo{volume}{2}}, 
  \bibinfo{pages}{20}.

\bibitem[{\citenamefont{Kimura et~al.}(2003)\citenamefont{Kimura, Goto,
  Shintani, Ishizaka, Arima and Tokura}}]{MF1}
\bibinfo{author}{\bibfnamefont{T.}~\bibnamefont{Kimura}},
  \bibinfo{author}{\bibfnamefont{T.}~\bibnamefont{Goto}},
  \bibinfo{author}{\bibfnamefont{H.}~\bibnamefont{Shintani}},
  \bibinfo{author}{\bibfnamefont{K.}~\bibnamefont{Ishizaka}},
  \bibinfo{author}{\bibfnamefont{T.}~\bibnamefont{Arima}} \bibnamefont{and}
  \bibinfo{author}{\bibfnamefont{Y.}~\bibnamefont{Tokura}},
  \bibinfo{journal}{Nature (London)}, \bibinfo{year}{2003}, \textbf{\bibinfo{volume}{426}},
  \bibinfo{pages}{55}.

\bibitem[{\citenamefont{Hur et~al.}(2004)\citenamefont{Hur, Park, Sharma, Ahn,
  Guha and Cheong}}]{MF2}
\bibinfo{author}{\bibfnamefont{N.}~\bibnamefont{Hur}},
  \bibinfo{author}{\bibfnamefont{S.}~\bibnamefont{Park}},
  \bibinfo{author}{\bibfnamefont{P.~A.} \bibnamefont{Sharma}},
  \bibinfo{author}{\bibfnamefont{J.~S.} \bibnamefont{Ahn}},
  \bibinfo{author}{\bibfnamefont{S.}~\bibnamefont{Guha}} \bibnamefont{and}
  \bibinfo{author}{\bibfnamefont{S.-W.} \bibnamefont{Cheong}},
  \bibinfo{journal}{Nature (London)}, \bibinfo{year}{2004}, \textbf{\bibinfo{volume}{429}},
  \bibinfo{pages}{392}.

\bibitem[{\citenamefont{Cheong and Mostovoy}(2007)}]{MF3}
\bibinfo{author}{\bibfnamefont{S.-W.} \bibnamefont{Cheong}} \bibnamefont{and}
  \bibinfo{author}{\bibfnamefont{M.}~\bibnamefont{Mostovoy}},
  \bibinfo{journal}{Nat. Mat.}, \bibinfo{year}{2007}, \textbf{\bibinfo{volume}{6}},
  \bibinfo{pages}{13}.

\bibitem[{\citenamefont{Wang et~al.}(2005)
\citenamefont{Wang, Neaton, Zheng, Nagarajan, Ogale, Liu, Viehland, Vaithyanathan, Schlom, Waghmareet~al.}}]{bfo}
\bibinfo{author}{\bibfnamefont{J.}~\bibnamefont{Wang}},
  \bibinfo{author}{\bibfnamefont{J.}~\bibnamefont{Neaton}},
  \bibinfo{author}{\bibfnamefont{H.}~\bibnamefont{Zheng}},
  \bibinfo{author}{\bibfnamefont{V.}~\bibnamefont{Nagarajan}},
  \bibinfo{author}{\bibfnamefont{S.~B.} \bibnamefont{Ogale}},
  \bibinfo{author}{\bibfnamefont{B.}~\bibnamefont{Liu}},
  \bibinfo{author}{\bibfnamefont{D.}~\bibnamefont{Viehland}},
  \bibinfo{author}{\bibfnamefont{V.}~\bibnamefont{Vaithyanathan}},
  \bibinfo{author}{\bibfnamefont{D.~G.} \bibnamefont{Schlom}},
  \bibinfo{author}{\bibfnamefont{U.~V.} \bibnamefont{Waghmare}},
  \bibnamefont{et~al.}, \bibinfo{journal}{Science}, \bibinfo{year}{2005},
  \textbf{\bibinfo{volume}{307}}, \bibinfo{pages}{1203b}.

\bibitem[{\citenamefont{Fiebig}(2005)}]{MF4}
\bibinfo{author}{\bibfnamefont{M.}~\bibnamefont{Fiebig}}, \bibinfo{journal}{J.
  Phys. D: Appl. Phys.}, \bibinfo{year}{2005}, \textbf{\bibinfo{volume}{38}}, 
  \bibinfo{pages}{R123}.

\bibitem[{\citenamefont{Spaldin and Fiebig}(2005)}]{MF7}
\bibinfo{author}{\bibfnamefont{N.~A.} \bibnamefont{Spaldin}} \bibnamefont{and}
  \bibinfo{author}{\bibfnamefont{M.}~\bibnamefont{Fiebig}},
  \bibinfo{journal}{Science}, \bibinfo{year}{2005}, \textbf{\bibinfo{volume}{309}},
  \bibinfo{pages}{391}.

\bibitem[{\citenamefont{K{\"u}mmel and Kronik}(2008)}]{failures}
\bibinfo{author}{\bibfnamefont{S.}~\bibnamefont{K{\"u}mmel}} \bibnamefont{and}
  \bibinfo{author}{\bibfnamefont{L.}~\bibnamefont{Kronik}},
  \bibinfo{journal}{Rev. Mod. Phys.}, \bibinfo{year}{2008}, \textbf{\bibinfo{volume}{80}},
  \bibinfo{pages}{3}.

\bibitem[{\citenamefont{Bachelet and Christensen}(1985)}]{MetallicGe}
\bibinfo{author}{\bibfnamefont{G.~B.} \bibnamefont{Bachelet}} \bibnamefont{and}
  \bibinfo{author}{\bibfnamefont{N.~E.} \bibnamefont{Christensen}},
  \bibinfo{journal}{Phys. Rev. B}, \bibinfo{year}{1985}, \textbf{\bibinfo{volume}{31}},
  \bibinfo{pages}{879}.

\bibitem[{\citenamefont{Filippetti and Hill}(2002)}]{hexYMnO3}
\bibinfo{author}{\bibfnamefont{A.}~\bibnamefont{Filippetti}} \bibnamefont{and}
  \bibinfo{author}{\bibfnamefont{N.~A.} \bibnamefont{Hill}},
  \bibinfo{journal}{Phys. Rev. B}, \bibinfo{year}{2002}, \textbf{\bibinfo{volume}{65}},
  \bibinfo{pages}{195120}.

\bibitem[{\citenamefont{Aulbur et~al.}(1999)\citenamefont{Aulbur, J{\'o}nsson
  and Wilkins}}]{ReviewGW1}
\bibinfo{author}{\bibfnamefont{W.~G.} \bibnamefont{Aulbur}},
  \bibinfo{author}{\bibfnamefont{L.}~\bibnamefont{J{\'o}nsson}}
  \bibnamefont{and} \bibinfo{author}{\bibfnamefont{J.~W.}
  \bibnamefont{Wilkins}}, \bibinfo{journal}{Solid State Physics}, \bibinfo{year}{1999}, 
  \textbf{\bibinfo{volume}{54}}, \bibinfo{pages}{1}.

\bibitem[{\citenamefont{Aryasetiawan and Gunnarsson}(1998)}]{ReviewGW2}
\bibinfo{author}{\bibfnamefont{F.}~\bibnamefont{Aryasetiawan}}
  \bibnamefont{and}
  \bibinfo{author}{\bibfnamefont{O.}~\bibnamefont{Gunnarsson}},
  \bibinfo{journal}{Rep. Prog. Phys.}, \bibinfo{year}{1998}, \textbf{\bibinfo{volume}{61}},
  \bibinfo{pages}{237}.

\bibitem[{\citenamefont{Anisimov et~al.}(1997)\citenamefont{Anisimov,
  Aryasetiawan and Lichtenstein}}]{ldau1}
\bibinfo{author}{\bibfnamefont{V.~I.} \bibnamefont{Anisimov}},
  \bibinfo{author}{\bibfnamefont{F.}~\bibnamefont{Aryasetiawan}}
  \bibnamefont{and} \bibinfo{author}{\bibfnamefont{A.~I.}
  \bibnamefont{Lichtenstein}}, \bibinfo{journal}{J. Phys.: Condens. Matter}, 
  \bibinfo{year}{1997},
  \textbf{\bibinfo{volume}{9}}, \bibinfo{pages}{767}.

\bibitem[{\citenamefont{Anisimov et~al.}(1991)\citenamefont{Anisimov, Zaanen
  and Andersen}}]{ldau2}
\bibinfo{author}{\bibfnamefont{V.~I.} \bibnamefont{Anisimov}},
  \bibinfo{author}{\bibfnamefont{J.}~\bibnamefont{Zaanen}} \bibnamefont{and}
  \bibinfo{author}{\bibfnamefont{O.~K.} \bibnamefont{Andersen}},
  \bibinfo{journal}{Phys. Rev. B}, \bibinfo{year}{1991}, \textbf{\bibinfo{volume}{44}},
  \bibinfo{pages}{943}.

\bibitem[{\citenamefont{Solovyev et~al.}(1994)\citenamefont{Solovyev,
  Dederichs and Anisimov}}]{ldau3}
\bibinfo{author}{\bibfnamefont{I.~V.} \bibnamefont{Solovyev}},
  \bibinfo{author}{\bibfnamefont{P.~H.} \bibnamefont{Dederichs}}
  \bibnamefont{and} \bibinfo{author}{\bibfnamefont{V.~I.}
  \bibnamefont{Anisimov}}, \bibinfo{journal}{Phys. Rev. B}, \bibinfo{year}{1994}, 
  \textbf{\bibinfo{volume}{50}}, \bibinfo{pages}{16861}.

\bibitem[{\citenamefont{Perdew and Zunger}(1981)}]{sic1}
\bibinfo{author}{\bibfnamefont{J.~P.} \bibnamefont{Perdew}} \bibnamefont{and}
  \bibinfo{author}{\bibfnamefont{A.}~\bibnamefont{Zunger}},
  \bibinfo{journal}{Phys. Rev. B}, \bibinfo{year}{1981}, \textbf{\bibinfo{volume}{23}},
  \bibinfo{pages}{5048}.

\bibitem[{\citenamefont{Filippetti and Spaldin}(2003)}]{sic2}
\bibinfo{author}{\bibfnamefont{A.}~\bibnamefont{Filippetti}} \bibnamefont{and}
  \bibinfo{author}{\bibfnamefont{N.~A.} \bibnamefont{Spaldin}},
  \bibinfo{journal}{Phys. Rev. B}, \bibinfo{year}{2003}, \textbf{\bibinfo{volume}{67}},
  \bibinfo{pages}{125109}.

\bibitem[{\citenamefont{Droghetti et~al.}(2008)\citenamefont{Droghetti,
  Pemmaraju and Sanvito}}]{sic3}
\bibinfo{author}{\bibfnamefont{A.}~\bibnamefont{Droghetti}},
  \bibinfo{author}{\bibfnamefont{C.~D.} \bibnamefont{Pemmaraju}}
  \bibnamefont{and} \bibinfo{author}{\bibfnamefont{S.}~\bibnamefont{Sanvito}},
  \bibinfo{journal}{Phys. Rev. B}, \bibinfo{year}{2008}, \textbf{\bibinfo{volume}{78}},
  \bibinfo{pages}{140404(R)}.

\bibitem[{\citenamefont{Rohrbach et~al.}(2004)\citenamefont{Rohrbach, Hafner
  and Kresse}}]{ldau.kresse}
\bibinfo{author}{\bibfnamefont{A.}~\bibnamefont{Rohrbach}},
  \bibinfo{author}{\bibfnamefont{J.}~\bibnamefont{Hafner}} \bibnamefont{and}
  \bibinfo{author}{\bibfnamefont{G.}~\bibnamefont{Kresse}},
  \bibinfo{journal}{Phys. Rev. B}, \bibinfo{year}{2004}, \textbf{\bibinfo{volume}{69}},
  \bibinfo{pages}{075413}.

\bibitem[{\citenamefont{Kornev et~al.}(2007)\citenamefont{Kornev, Lisenkov,
  Haumont, Dkhil and Bellaiche}}]{Ueff}
\bibinfo{author}{\bibfnamefont{I.~A.} \bibnamefont{Kornev}},
  \bibinfo{author}{\bibfnamefont{S.}~\bibnamefont{Lisenkov}},
  \bibinfo{author}{\bibfnamefont{R.}~\bibnamefont{Haumont}},
  \bibinfo{author}{\bibfnamefont{B.}~\bibnamefont{Dkhil}} \bibnamefont{and}
  \bibinfo{author}{\bibfnamefont{L.}~\bibnamefont{Bellaiche}},
  \bibinfo{journal}{Phys. Rev. Lett.}, \bibinfo{year}{2007}, \textbf{\bibinfo{volume}{99}},
  \bibinfo{pages}{227602}.

\bibitem[{\citenamefont{Svane and Gunnarsson}(1990)}]{Svane}
\bibinfo{author}{\bibfnamefont{A.}~\bibnamefont{Svane}} \bibnamefont{and}
  \bibinfo{author}{\bibfnamefont{O.}~\bibnamefont{Gunnarsson}},
  \bibinfo{journal}{Phys. Rev. Lett.}, \bibinfo{year}{1990}, \textbf{\bibinfo{volume}{65}},
  \bibinfo{pages}{1148}.

\bibitem[{\citenamefont{Pemmaraju et~al.}(2007)\citenamefont{Pemmaraju, Archer,
  S\'{a}nchez-Portal and Sanvito}}]{AtomicSic}
\bibinfo{author}{\bibfnamefont{C.~D.} \bibnamefont{Pemmaraju}},
  \bibinfo{author}{\bibfnamefont{T.}~\bibnamefont{Archer}},
  \bibinfo{author}{\bibfnamefont{D.}~\bibnamefont{S\'{a}nchez-Portal}}
  \bibnamefont{and} \bibinfo{author}{\bibfnamefont{S.}~\bibnamefont{Sanvito}},
  \bibinfo{journal}{Phys. Rev. B}, \bibinfo{year}{2007}, \textbf{\bibinfo{volume}{75}},
  \bibinfo{pages}{045101}.

\bibitem[{\citenamefont{Stengel and Spaldin}(2008)}]{Stengel}
\bibinfo{author}{\bibfnamefont{M.}~\bibnamefont{Stengel}} \bibnamefont{and}
  \bibinfo{author}{\bibfnamefont{N.~A.} \bibnamefont{Spaldin}},
  \bibinfo{journal}{Phys. Rev. B}, \bibinfo{year}{2008}, \textbf{\bibinfo{volume}{77}},
  \bibinfo{pages}{155106}.

\bibitem[{\citenamefont{Stephens et~al.}(1994)\citenamefont{Stephens, Devlin,
  Chabalowski and Frisch}}]{hybrid2}
\bibinfo{author}{\bibfnamefont{P.~J.} \bibnamefont{Stephens}},
  \bibinfo{author}{\bibfnamefont{F.~J.} \bibnamefont{Devlin}},
  \bibinfo{author}{\bibfnamefont{C.}~\bibnamefont{Chabalowski}}
  \bibnamefont{and} \bibinfo{author}{\bibfnamefont{M.~J.}
  \bibnamefont{Frisch}}, \bibinfo{journal}{J. Phys. Chem.}, \bibinfo{year}{1994}, 
  \textbf{\bibinfo{volume}{98}}, \bibinfo{pages}{11623}.

\bibitem[{\citenamefont{Adamo and Barone}(1999)}]{hybrid3}
\bibinfo{author}{\bibfnamefont{C.}~\bibnamefont{Adamo}} \bibnamefont{and}
  \bibinfo{author}{\bibfnamefont{V.}~\bibnamefont{Barone}},
  \bibinfo{journal}{J. Chem. Phys.}, \bibinfo{year}{1999}, \textbf{\bibinfo{volume}{110}},
  \bibinfo{pages}{6158}.

\bibitem[{\citenamefont{Ernzerhof and Scuseria}(1999)}]{hybrid4}
\bibinfo{author}{\bibfnamefont{M.}~\bibnamefont{Ernzerhof}} \bibnamefont{and}
  \bibinfo{author}{\bibfnamefont{G.~E.} \bibnamefont{Scuseria}},
  \bibinfo{journal}{J. Chem. Phys.}, \bibinfo{year}{1999}, \textbf{\bibinfo{volume}{110}},
  \bibinfo{pages}{5029}.

\bibitem[{\citenamefont{Krukau et~al.}(2006)\citenamefont{Krukau, Vydrov,
  Izmaylov and Scuseria}}]{hybrid5}
\bibinfo{author}{\bibfnamefont{A.~V.} \bibnamefont{Krukau}},
  \bibinfo{author}{\bibfnamefont{O.~A.} \bibnamefont{Vydrov}},
  \bibinfo{author}{\bibfnamefont{A.~F.} \bibnamefont{Izmaylov}}
  \bibnamefont{and} \bibinfo{author}{\bibfnamefont{G.~E.}
  \bibnamefont{Scuseria}}, \bibinfo{journal}{J. Chem. Phys.}, \bibinfo{year}{2006},
  \textbf{\bibinfo{volume}{125}}, \bibinfo{pages}{224106}.

\bibitem[{\citenamefont{Muscat et~al.}(2001)\citenamefont{Muscat, Wander and
  Harrison}}]{hybrid6}
\bibinfo{author}{\bibfnamefont{J.}~\bibnamefont{Muscat}},
  \bibinfo{author}{\bibfnamefont{A.}~\bibnamefont{Wander}} \bibnamefont{and}
  \bibinfo{author}{\bibfnamefont{N.~M.} \bibnamefont{Harrison}},
  \bibinfo{journal}{Chem. Phys. Lett.}, \bibinfo{year}{2001}, \textbf{\bibinfo{volume}{342}},
  \bibinfo{pages}{397}.

\bibitem[{\citenamefont{Heyd et~al.}(2005)\citenamefont{Heyd, Peralta,
  Scuseria and Martin}}]{hybrid7}
\bibinfo{author}{\bibfnamefont{J.}~\bibnamefont{Heyd}},
  \bibinfo{author}{\bibfnamefont{J.~E.} \bibnamefont{Peralta}},
  \bibinfo{author}{\bibfnamefont{G.~E.} \bibnamefont{Scuseria}}
  \bibnamefont{and} \bibinfo{author}{\bibfnamefont{R.~L.}
  \bibnamefont{Martin}}, \bibinfo{journal}{J. Chem. Phys.}, \bibinfo{year}{2005},
  \textbf{\bibinfo{volume}{123}}, \bibinfo{pages}{174101}.

\bibitem[{\citenamefont{Paier et~al.}(2006)\citenamefont{Paier, Marsman,
  Hummer, Kresse, Gerber and \'{A}ngy\'{a}n}}]{hybrid7.5}
\bibinfo{author}{\bibfnamefont{J.}~\bibnamefont{Paier}},
  \bibinfo{author}{\bibfnamefont{M.}~\bibnamefont{Marsman}},
  \bibinfo{author}{\bibfnamefont{K.}~\bibnamefont{Hummer}},
  \bibinfo{author}{\bibfnamefont{G.}~\bibnamefont{Kresse}},
  \bibinfo{author}{\bibfnamefont{I.~C.} \bibnamefont{Gerber}}
  \bibnamefont{and} \bibinfo{author}{\bibfnamefont{J.~A.}
  \bibnamefont{\'{A}ngy\'{a}n}}, \bibinfo{journal}{J. Chem. Phys.}, \bibinfo{year}{2006},
  \textbf{\bibinfo{volume}{124}}, \bibinfo{pages}{154709}.

\bibitem[{\citenamefont{Hay et~al.}(2006)\citenamefont{Hay, Martin, Uddin and
  Scuseria}}]{itinerant1}
\bibinfo{author}{\bibfnamefont{P.~J.} \bibnamefont{Hay}},
  \bibinfo{author}{\bibfnamefont{R.~L.} \bibnamefont{Martin}},
  \bibinfo{author}{\bibfnamefont{J.}~\bibnamefont{Uddin}} \bibnamefont{and}
  \bibinfo{author}{\bibfnamefont{G.~E.} \bibnamefont{Scuseria}},
  \bibinfo{journal}{J. Chem. Phys.}, \bibinfo{year}{2006}, \textbf{\bibinfo{volume}{125}},
  \bibinfo{pages}{034712}.

\bibitem[{\citenamefont{Prodan et~al.}(2006)\citenamefont{Prodan, Scuseria and
  Martin}}]{itinerant2}
\bibinfo{author}{\bibfnamefont{I.~D.} \bibnamefont{Prodan}},
  \bibinfo{author}{\bibfnamefont{G.~E.} \bibnamefont{Scuseria}}
  \bibnamefont{and} \bibinfo{author}{\bibfnamefont{R.~L.}
  \bibnamefont{Martin}}, \bibinfo{journal}{Phys. Rev. B}, \bibinfo{year}{2006},
  \textbf{\bibinfo{volume}{73}}, \bibinfo{pages}{045104}.

\bibitem[{\citenamefont{Prodan et~al.}(2007)\citenamefont{Prodan, Scuseria and
  Martin}}]{itinerant3}
\bibinfo{author}{\bibfnamefont{I.~D.} \bibnamefont{Prodan}},
  \bibinfo{author}{\bibfnamefont{G.~E.} \bibnamefont{Scuseria}}
  \bibnamefont{and} \bibinfo{author}{\bibfnamefont{R.~L.}
  \bibnamefont{Martin}}, \bibinfo{journal}{Phys. Rev. B}, \bibinfo{year}{2007},
  \textbf{\bibinfo{volume}{76}}, \bibinfo{pages}{033101}.

\bibitem[{\citenamefont{Marsman et~al.}(2008)\citenamefont{Marsman, Paier,
  Stroppa and Kresse}}]{itinerant4}
\bibinfo{author}{\bibfnamefont{M.}~\bibnamefont{Marsman}},
  \bibinfo{author}{\bibfnamefont{J.}~\bibnamefont{Paier}},
  \bibinfo{author}{\bibfnamefont{A.}~\bibnamefont{Stroppa}} \bibnamefont{and}
  \bibinfo{author}{\bibfnamefont{G.}~\bibnamefont{Kresse}},
  \bibinfo{journal}{J. Phys.: Condens. Matter.}, \bibinfo{year}{2008}, \textbf{\bibinfo{volume}{20}},
  \bibinfo{pages}{064201}.

\bibitem[{\citenamefont{Stroppa et~al.}(2007)\citenamefont{Stroppa,
  Termentzidis, Paier, Kresse and Hafner}}]{Stroppa1}
\bibinfo{author}{\bibfnamefont{A.}~\bibnamefont{Stroppa}},
  \bibinfo{author}{\bibfnamefont{K.}~\bibnamefont{Termentzidis}},
  \bibinfo{author}{\bibfnamefont{J.}~\bibnamefont{Paier}},
  \bibinfo{author}{\bibfnamefont{G.}~\bibnamefont{Kresse}} \bibnamefont{and}
  \bibinfo{author}{\bibfnamefont{J.}~\bibnamefont{Hafner}},
  \bibinfo{journal}{Phys. Rev. B}, \bibinfo{year}{2007}, \textbf{\bibinfo{volume}{76}},
  \bibinfo{pages}{195440}.

\bibitem[{\citenamefont{Stroppa and Kresse}(2008)}]{Stroppa2}
\bibinfo{author}{\bibfnamefont{A.}~\bibnamefont{Stroppa}} \bibnamefont{and}
  \bibinfo{author}{\bibfnamefont{G.}~\bibnamefont{Kresse}},
  \bibinfo{journal}{New J. Phys.}, \bibinfo{year}{2008}, \textbf{\bibinfo{volume}{10}},
  \bibinfo{pages}{063020}.

\bibitem[{\citenamefont{Stroppa and Kresse}(2009)}]{Stroppa3}
\bibinfo{author}{\bibfnamefont{A.}~\bibnamefont{Stroppa}} \bibnamefont{and}
  \bibinfo{author}{\bibfnamefont{G.}~\bibnamefont{Kresse}},
  \bibinfo{journal}{Phys. Rev. B (R)}, \bibinfo{year}{2009}, \textbf{\bibinfo{volume}{79}},
  \bibinfo{pages}{201201(R)}.

\bibitem[{\citenamefont{Becke}(1993)}]{hybrid11}
\bibinfo{author}{\bibfnamefont{A.~D.} \bibnamefont{Becke}},
  \bibinfo{journal}{J. Chem. Phys.}, \bibinfo{year}{1993}, \textbf{\bibinfo{volume}{98}},
  \bibinfo{pages}{1372}.

\bibitem[{\citenamefont{Ernzerhof et~al.}(1996)\citenamefont{Ernzerhof, Burke
  and Perdew}}]{hybrid12}
\bibinfo{author}{\bibfnamefont{M.}~\bibnamefont{Ernzerhof}},
  \bibinfo{author}{\bibfnamefont{K.}~\bibnamefont{Burke}} \bibnamefont{and}
  \bibinfo{author}{\bibfnamefont{J.~P.} \bibnamefont{Perdew}},
  \bibinfo{journal}{J. Chem. Phys.}, \bibinfo{year}{1996}, \textbf{\bibinfo{volume}{105}},
  \bibinfo{pages}{9982}.

\bibitem[{\citenamefont{Janesko et~al.}(2009)\citenamefont{Janesko, Henderson
  and Scuseria}}]{ScuseriaReview}
\bibinfo{author}{\bibfnamefont{B.~J.} \bibnamefont{Janesko}},
  \bibinfo{author}{\bibfnamefont{T.~M.} \bibnamefont{Henderson}}
  \bibnamefont{and} \bibinfo{author}{\bibfnamefont{G.}~\bibnamefont{Scuseria}},
  \bibinfo{journal}{Phys. Chem. Chem. Phys.}, \bibinfo{year}{2009}, \textbf{\bibinfo{volume}{11}},
  \bibinfo{pages}{443}.

\bibitem[{\citenamefont{Heyd et~al.}(2003)\citenamefont{Heyd, Scuseria and
  Ernzerhof}}]{hsedef}
\bibinfo{author}{\bibfnamefont{J.}~\bibnamefont{Heyd}},
  \bibinfo{author}{\bibfnamefont{G.~E.} \bibnamefont{Scuseria}}
  \bibnamefont{and}
  \bibinfo{author}{\bibfnamefont{M.}~\bibnamefont{Ernzerhof}},
  \bibinfo{journal}{J. Chem. Phys.}, \bibinfo{year}{2003}, \textbf{\bibinfo{volume}{118}},
  \bibinfo{pages}{8207}.

\bibitem[{\citenamefont{Heyd et~al.}(2006)\citenamefont{Heyd, Scuseria and
  Ernzerhof}}]{hsedef1}
\bibinfo{author}{\bibfnamefont{J.}~\bibnamefont{Heyd}},
  \bibinfo{author}{\bibfnamefont{G.~E.} \bibnamefont{Scuseria}}
  \bibnamefont{and}
  \bibinfo{author}{\bibfnamefont{M.}~\bibnamefont{Ernzerhof}},
  \bibinfo{journal}{J. Chem. Phys.}, \bibinfo{year}{2006}, \textbf{\bibinfo{volume}{124}},
  \bibinfo{pages}{219906}.

\bibitem[{\citenamefont{Wahl et~al.}(2008)\citenamefont{Wahl, Vogtenhuber and
  Kresse}}]{Roman}
\bibinfo{author}{\bibfnamefont{R.}~\bibnamefont{Wahl}},
  \bibinfo{author}{\bibfnamefont{D.}~\bibnamefont{Vogtenhuber}}
  \bibnamefont{and} \bibinfo{author}{\bibfnamefont{G.}~\bibnamefont{Kresse}},
  \bibinfo{journal}{Phys. Rev. B}, \bibinfo{year}{2008}, \textbf{\bibinfo{volume}{78}},
  \bibinfo{pages}{104116}.

\bibitem[{\citenamefont{Perdew et~al.}(1996)\citenamefont{Perdew, Burke and
  Ernzerhof}}]{pbe}
\bibinfo{author}{\bibfnamefont{J.~P.}~\bibnamefont{Perdew}},
  \bibinfo{author}{\bibfnamefont{K.}~\bibnamefont{Burke}} \bibnamefont{and}
  \bibinfo{author}{\bibfnamefont{M.}~\bibnamefont{Ernzerhof}},
  \bibinfo{journal}{Phys. Rev. Lett.}, \bibinfo{year}{1996}, \textbf{\bibinfo{volume}{77}},
  \bibinfo{pages}{3865}.

\bibitem[{\citenamefont{Bilc et~al.}(2008)\citenamefont{Bilc, Orlando, Shaltaf,
  Rignanese, {\'I}{\~n}iguez and Ghosez}}]{Bilc}
\bibinfo{author}{\bibfnamefont{D.~I.} \bibnamefont{Bilc}},
  \bibinfo{author}{\bibfnamefont{R.}~\bibnamefont{Orlando}},
  \bibinfo{author}{\bibfnamefont{R.}~\bibnamefont{Shaltaf}},
  \bibinfo{author}{\bibfnamefont{G.-M.} \bibnamefont{Rignanese}},
  \bibinfo{author}{\bibfnamefont{J.}~\bibnamefont{{\'I}{\~n}iguez}}
  \bibnamefont{and} \bibinfo{author}{\bibfnamefont{P.}~\bibnamefont{Ghosez}},
  \bibinfo{journal}{Phys. Rev. B}, \bibinfo{year}{2008}, \textbf{\bibinfo{volume}{77}},
  \bibinfo{pages}{165107}.

\bibitem[{\citenamefont{Goffinet et~al.}(2009)\citenamefont{Goffinet, Bilc and
  Ghosez}}]{Goffinet}
\bibinfo{author}{\bibfnamefont{M.}~\bibnamefont{Goffinet}},
  \bibinfo{author}{\bibfnamefont{P.}~\bibnamefont{Hermet}}, 
  \bibinfo{author}{\bibfnamefont{D.~I.} \bibnamefont{Bilc}}
  \bibnamefont{and} \bibinfo{author}{\bibfnamefont{P.}~\bibnamefont{Ghosez}},
  \bibinfo{journal}{Phys. Rev. B}, \bibinfo{year}{2009}, \textbf{\bibinfo{volume}{79}},
  \bibinfo{pages}{014403}.

\bibitem[{\citenamefont{Wu and Cohen}(2006)}]{WC}
\bibinfo{author}{\bibfnamefont{Z.}~\bibnamefont{Wu}} \bibnamefont{and}
  \bibinfo{author}{\bibfnamefont{R.~E.} \bibnamefont{Cohen}},
  \bibinfo{journal}{Phys. Rev. B}, \bibinfo{year}{2006}, \textbf{\bibinfo{volume}{73}},
  \bibinfo{pages}{235116}.

\bibitem[{\citenamefont{Heifets et~al.}(2006)\citenamefont{Heifets, Kotomin
  and Trepakov}}]{B1WC}
\bibinfo{author}{\bibfnamefont{E.}~\bibnamefont{Heifets}},
  \bibinfo{author}{\bibfnamefont{E.}~\bibnamefont{Kotomin}} \bibnamefont{and}
  \bibinfo{author}{\bibfnamefont{V.~A.} \bibnamefont{Trepakov}},
  \bibinfo{journal}{J. Phys.: Condens. Matter}, \bibinfo{year}{2006}, \textbf{\bibinfo{volume}{18}},
  \bibinfo{pages}{4845}.

\bibitem[{\citenamefont{Picozzi et~al.}(2007)\citenamefont{Picozzi, Yamauchi,
  Sanyal, Sergienko and Dagotto}}]{HMO1}
\bibinfo{author}{\bibfnamefont{S.}~\bibnamefont{Picozzi}},
  \bibinfo{author}{\bibfnamefont{K.}~\bibnamefont{Yamauchi}},
  \bibinfo{author}{\bibfnamefont{B.}~\bibnamefont{Sanyal}},
  \bibinfo{author}{\bibfnamefont{I.~A.} \bibnamefont{Sergienko}}
  \bibnamefont{and} \bibinfo{author}{\bibfnamefont{E.}~\bibnamefont{Dagotto}},
  \bibinfo{journal}{Phys. Rev. Lett.}, \bibinfo{year}{2007}, \textbf{\bibinfo{volume}{99}},
  \bibinfo{pages}{227201}.

\bibitem[{\citenamefont{Munoz et~al.}(2001)\citenamefont{Munoz, Cas\'{a}is,
  Alonso, Mart\'{i}nez-Lope, Mart\'{i}nez and Fern\'{a}ndez-D\'{i}az}}]{HMO2}
\bibinfo{author}{\bibfnamefont{A.}~\bibnamefont{Munoz}},
  \bibinfo{author}{\bibfnamefont{M.}~\bibnamefont{Cas\'{a}is}},
  \bibinfo{author}{\bibfnamefont{J.~A.} \bibnamefont{Alonso}},
  \bibinfo{author}{\bibfnamefont{M.~J.} \bibnamefont{Mart\'{i}nez-Lope}},
  \bibinfo{author}{\bibfnamefont{J.~L.} \bibnamefont{Mart\'{i}nez}}
  \bibnamefont{and}
  \bibinfo{author}{\bibfnamefont{M.}~\bibnamefont{Fern\'{a}ndez-D\'{i}az}},
  \bibinfo{journal}{Inorg. Chem.}, \bibinfo{year}{2001}, \textbf{\bibinfo{volume}{40}},
  \bibinfo{pages}{1020}.

\bibitem[{\citenamefont{Kresse and Furthm{\"u}ller}(1996)}]{vasp1}
\bibinfo{author}{\bibfnamefont{G.}~\bibnamefont{Kresse}} \bibnamefont{and}
  \bibinfo{author}{\bibfnamefont{J.}~\bibnamefont{Furthm{\"u}ller}},
  \bibinfo{journal}{Phys. Rev. B}, \bibinfo{year}{1996}, \textbf{\bibinfo{volume}{54}},
  \bibinfo{pages}{11169}.

\bibitem[{\citenamefont{Bl{\"o}chl}(1994)}]{paw1}
\bibinfo{author}{\bibfnamefont{P.~E.} \bibnamefont{Bl{\"o}chl}},
  \bibinfo{journal}{Phys. Rev. B}, \bibinfo{year}{1994}, \textbf{\bibinfo{volume}{50}},
  \bibinfo{pages}{17953}.

\bibitem[{\citenamefont{Kresse and Joubert}(1999)}]{paw2}
\bibinfo{author}{\bibfnamefont{G.}~\bibnamefont{Kresse}} \bibnamefont{and}
  \bibinfo{author}{\bibfnamefont{D.}~\bibnamefont{Joubert}},
  \bibinfo{journal}{Phys. Rev. B}, \bibinfo{year}{1999}, \textbf{\bibinfo{volume}{59}},
  \bibinfo{pages}{1758}.

\bibitem[{\citenamefont{Boysen and Altorfer}(1994)}]{paraelectric}
\bibinfo{author}{\bibfnamefont{H.}~\bibnamefont{Boysen}} \bibnamefont{and}
  \bibinfo{author}{\bibfnamefont{F.}~\bibnamefont{Altorfer}},
  \bibinfo{journal}{Acta Cryst. B}, \bibinfo{year}{1994}, \textbf{\bibinfo{volume}{50}},
  \bibinfo{pages}{405}.

\bibitem[{\citenamefont{Neaton et~al.}(2005)\citenamefont{Neaton, Ederer,
  Waghmare, Spaldin and Rabe}}]{Neaton}
\bibinfo{author}{\bibfnamefont{J.~B.} \bibnamefont{Neaton}},
  \bibinfo{author}{\bibfnamefont{C.}~\bibnamefont{Ederer}},
  \bibinfo{author}{\bibfnamefont{U.~V.} \bibnamefont{Waghmare}},
  \bibinfo{author}{\bibfnamefont{N.~A.} \bibnamefont{Spaldin}}
  \bibnamefont{and} \bibinfo{author}{\bibfnamefont{K.~M.} \bibnamefont{Rabe}},
  \bibinfo{journal}{Phys. Rev. B}, \bibinfo{year}{2005}, \textbf{\bibinfo{volume}{71}},
  \bibinfo{pages}{014113}.

\bibitem[{\citenamefont{King-Smith and Vanderbilt}(1993)}]{berry1}
\bibinfo{author}{\bibfnamefont{R.~D.} \bibnamefont{King-Smith}}
  \bibnamefont{and}
  \bibinfo{author}{\bibfnamefont{D.}~\bibnamefont{Vanderbilt}},
  \bibinfo{journal}{Phys. Rev. B}, \bibinfo{year}{1993}, \textbf{\bibinfo{volume}{47}},
  \bibinfo{pages}{1651}.

\bibitem[{\citenamefont{Vanderbilt and King-Smith}(1993)}]{berry2}
\bibinfo{author}{\bibfnamefont{D.}~\bibnamefont{Vanderbilt}} \bibnamefont{and}
  \bibinfo{author}{\bibfnamefont{R.~D.} \bibnamefont{King-Smith}},
  \bibinfo{journal}{Phys. Rev. B}, \bibinfo{year}{1993}, \textbf{\bibinfo{volume}{48}},
  \bibinfo{pages}{4442}.

\bibitem[{\citenamefont{Resta}(1994)}]{berry3}
\bibinfo{author}{\bibfnamefont{R.}~\bibnamefont{Resta}}, \bibinfo{journal}{Rev.
  Mod. Phys.}, \bibinfo{year}{1994}, \textbf{\bibinfo{volume}{66}}, \bibinfo{pages}{899}.

\bibitem[{\citenamefont{Hybertsen and Louie}(1986)}]{GW}
\bibinfo{author}{\bibfnamefont{M.~S.} \bibnamefont{Hybertsen}}
  \bibnamefont{and} \bibinfo{author}{\bibfnamefont{S.~G.} \bibnamefont{Louie}},
  \bibinfo{journal}{Phys. Rev. B}, \bibinfo{year}{1986}, \textbf{\bibinfo{volume}{34}},
  \bibinfo{pages}{5390}.

\bibitem[{\citenamefont{Fuchs et~al.}(2007)\citenamefont{Fuchs, Furthmuller,
  Bechstedt, Shishkin and Kresse}}]{GW1}
\bibinfo{author}{\bibfnamefont{F.}~\bibnamefont{Fuchs}},
  \bibinfo{author}{\bibfnamefont{J.}~\bibnamefont{Furthmuller}},
  \bibinfo{author}{\bibfnamefont{F.}~\bibnamefont{Bechstedt}},
  \bibinfo{author}{\bibfnamefont{M.}~\bibnamefont{Shishkin}} \bibnamefont{and}
  \bibinfo{author}{\bibfnamefont{G.}~\bibnamefont{Kresse}},
  \bibinfo{journal}{Phys. Rev. B}, \bibinfo{year}{2007}, \textbf{\bibinfo{volume}{76}},
  \bibinfo{pages}{115109}.

\bibitem[{\citenamefont{Shishkin and Kresse}(2006)}]{GW2}
\bibinfo{author}{\bibfnamefont{M.}~\bibnamefont{Shishkin}} \bibnamefont{and}
  \bibinfo{author}{\bibfnamefont{G.}~\bibnamefont{Kresse}},
  \bibinfo{journal}{Phys. Rev. B}, \bibinfo{year}{2006}, \textbf{\bibinfo{volume}{74}},
  \bibinfo{pages}{035101}.

\bibitem[{\citenamefont{Shishkin and Kresse}(2007)}]{GW3}
\bibinfo{author}{\bibfnamefont{M.}~\bibnamefont{Shishkin}} \bibnamefont{and}
  \bibinfo{author}{\bibfnamefont{G.}~\bibnamefont{Kresse}},
  \bibinfo{journal}{Phys. Rev. B}, \bibinfo{year}{2007}, \textbf{\bibinfo{volume}{75}},
  \bibinfo{pages}{235102}.

\bibitem[{\citenamefont{Shishkin et~al.}(2007)\citenamefont{Shishkin, Marsman
  and Kresse}}]{GW4}
\bibinfo{author}{\bibfnamefont{M.}~\bibnamefont{Shishkin}},
  \bibinfo{author}{\bibfnamefont{M.}~\bibnamefont{Marsman}} \bibnamefont{and}
  \bibinfo{author}{\bibfnamefont{G.}~\bibnamefont{Kresse}},
  \bibinfo{journal}{Phys. Rev. Lett.}, \bibinfo{year}{2007}, \textbf{\bibinfo{volume}{99}},
  \bibinfo{pages}{246403}.

\bibitem[{\citenamefont{Rinke et~al.}(2005)\citenamefont{Rinke, Qteish,
  Neugebauer, Freysoldt and Scheffler}}]{GW5}
\bibinfo{author}{\bibfnamefont{P.}~\bibnamefont{Rinke}},
  \bibinfo{author}{\bibfnamefont{A.}~\bibnamefont{Qteish}},
  \bibinfo{author}{\bibfnamefont{J.}~\bibnamefont{Neugebauer}},
  \bibinfo{author}{\bibfnamefont{C.}~\bibnamefont{Freysoldt}}
  \bibnamefont{and}
  \bibinfo{author}{\bibfnamefont{M.}~\bibnamefont{Scheffler}},
  \bibinfo{journal}{New J. Phys.}, \bibinfo{year}{2005}, \textbf{\bibinfo{volume}{7}},
  \bibinfo{pages}{125}.

\bibitem[{\citenamefont{Kubel and Schmid}(1990)}]{Kubel}
\bibinfo{author}{\bibfnamefont{F.}~\bibnamefont{Kubel}} \bibnamefont{and}
  \bibinfo{author}{\bibfnamefont{H.}~\bibnamefont{Schmid}},
  \bibinfo{journal}{Acta Crystallogr., Sect. B: Struct. Sci.},
   \bibinfo{year}{1990}, 
  \textbf{\bibinfo{volume}{B46}}, \bibinfo{pages}{698}.

\bibitem[{\citenamefont{Gavriliuk et~al.}(2008)\citenamefont{Gavriliuk,
  Struzhkin, Lyubutin, Ovchinnikov, Hu and Chow}}]{gap1}
\bibinfo{author}{\bibfnamefont{A.~G.} \bibnamefont{Gavriliuk}},
  \bibinfo{author}{\bibfnamefont{V.~V.} \bibnamefont{Struzhkin}},
  \bibinfo{author}{\bibfnamefont{I.~S.} \bibnamefont{Lyubutin}},
  \bibinfo{author}{\bibfnamefont{S.~G.} \bibnamefont{Ovchinnikov}},
  \bibinfo{author}{\bibfnamefont{M.~Y.} \bibnamefont{Hu}} \bibnamefont{and}
  \bibinfo{author}{\bibfnamefont{P.}~\bibnamefont{Chow}},
  \bibinfo{journal}{Phys. Rev. B}, \bibinfo{year}{2008},  \textbf{\bibinfo{volume}{77}},
  \bibinfo{pages}{155112}.

\bibitem[{\citenamefont{Palai et~al.}(2008)\citenamefont{Palai, Katiyar,
  Schmid, Tissot, Clark, Robertson, Redfern, Catalan and Scott}}]{gap2}
\bibinfo{author}{\bibfnamefont{R.}~\bibnamefont{Palai}},
  \bibinfo{author}{\bibfnamefont{R.~S.} \bibnamefont{Katiyar}},
  \bibinfo{author}{\bibfnamefont{H.}~\bibnamefont{Schmid}},
  \bibinfo{author}{\bibfnamefont{P.}~\bibnamefont{Tissot}},
  \bibinfo{author}{\bibfnamefont{S.~J.} \bibnamefont{Clark}},
  \bibinfo{author}{\bibfnamefont{J.}~\bibnamefont{Robertson}},
  \bibinfo{author}{\bibfnamefont{S.~A.~T.} \bibnamefont{Redfern}},
  \bibinfo{author}{\bibfnamefont{G.}~\bibnamefont{Catalan}} \bibnamefont{and}
  \bibinfo{author}{\bibfnamefont{J.~F.} \bibnamefont{Scott}},
  \bibinfo{journal}{Phys. Rev. B}, \bibinfo{year}{2008}, \textbf{\bibinfo{volume}{77}},
  \bibinfo{pages}{014110}.

\bibitem[{\citenamefont{Basu et~al.}(2008)\citenamefont{Basu, Martin, Chu,
  Gajek, Ramesh, Rai, Xu and Musfeldt}}]{gap3}
\bibinfo{author}{\bibfnamefont{S.~R.} \bibnamefont{Basu}},
  \bibinfo{author}{\bibfnamefont{L.~W.} \bibnamefont{Martin}},
  \bibinfo{author}{\bibfnamefont{Y.~H.} \bibnamefont{Chu}},
  \bibinfo{author}{\bibfnamefont{M.}~\bibnamefont{Gajek}},
  \bibinfo{author}{\bibfnamefont{R.}~\bibnamefont{Ramesh}},
  \bibinfo{author}{\bibfnamefont{R.~C.} \bibnamefont{Rai}},
  \bibinfo{author}{\bibfnamefont{X.}~\bibnamefont{Xu}} \bibnamefont{and}
  \bibinfo{author}{\bibfnamefont{J.~L.} \bibnamefont{Musfeldt}},
  \bibinfo{journal}{Appl. Phys. Lett.}, \bibinfo{year}{2008}, \textbf{\bibinfo{volume}{92}},
  \bibinfo{pages}{091905}.

\bibitem[{\citenamefont{Kumar et~al.}(2008)\citenamefont{Kumar, Rai, Podraza,
  Denev, Ramirez, Chu, Martin, Ihlefeld, Heeg, Schubert et~al.}}]{gap4}
\bibinfo{author}{\bibfnamefont{A.}~\bibnamefont{Kumar}},
  \bibinfo{author}{\bibfnamefont{R.~C.} \bibnamefont{Rai}},
  \bibinfo{author}{\bibfnamefont{N.~J.} \bibnamefont{Podraza}},
  \bibinfo{author}{\bibfnamefont{S.}~\bibnamefont{Denev}},
  \bibinfo{author}{\bibfnamefont{M.}~\bibnamefont{Ramirez}},
  \bibinfo{author}{\bibfnamefont{Y.~H.} \bibnamefont{Chu}},
  \bibinfo{author}{\bibfnamefont{L.~W.} \bibnamefont{Martin}},
  \bibinfo{author}{\bibfnamefont{J.}~\bibnamefont{Ihlefeld}},
  \bibinfo{author}{\bibfnamefont{T.}~\bibnamefont{Heeg}},
  \bibinfo{author}{\bibfnamefont{J.}~\bibnamefont{Schubert}},
  \bibnamefont{et~al.}, \bibinfo{journal}{Appl. Phys. Lett.}, \bibinfo{year}{2008},
  \textbf{\bibinfo{volume}{92}}, \bibinfo{pages}{121915}.

\bibitem[{\citenamefont{Kanai et~al.}(2003)\citenamefont{Kanai, Ohkoshi and
  Hashimoto}}]{Kanai}
\bibinfo{author}{\bibfnamefont{T.}~\bibnamefont{Kanai}},
  \bibinfo{author}{\bibfnamefont{S.}~\bibnamefont{Ohkoshi}} \bibnamefont{and}
  \bibinfo{author}{\bibfnamefont{K.}~\bibnamefont{Hashimoto}},
  \bibinfo{journal}{J. Phys. Chem. Solids}, \bibinfo{year}{2003}, \textbf{\bibinfo{volume}{64}},
  \bibinfo{pages}{391}.

\bibitem[{\citenamefont{Gao et~al.}(2006)\citenamefont{Gao, Yuan, Wang, Chen,
  Chen and Liu}}]{Gao}
\bibinfo{author}{\bibfnamefont{F.}~\bibnamefont{Gao}},
  \bibinfo{author}{\bibfnamefont{Y.}~\bibnamefont{Yuan}},
  \bibinfo{author}{\bibfnamefont{K.~F.} \bibnamefont{Wang}},
  \bibinfo{author}{\bibfnamefont{X.~Y.} \bibnamefont{Chen}},
  \bibinfo{author}{\bibfnamefont{F.}~\bibnamefont{Chen}} \bibnamefont{and}
  \bibinfo{author}{\bibfnamefont{J.~M.} \bibnamefont{Liu}},
  \bibinfo{journal}{Appl. Phys. Lett.}, \bibinfo{year}{2006}, \textbf{\bibinfo{volume}{89}},
  \bibinfo{pages}{102506}.

\bibitem[{\citenamefont{Higuchi et~al.}(2008)\citenamefont{Higuchi, Liu, Yao,
  Glans, Guo, Chang, Wu, Sakamoto, Itoh, Shimura et~al.}}]{Guo-exp-gap}
\bibinfo{author}{\bibfnamefont{T.}~\bibnamefont{Higuchi}},
  \bibinfo{author}{\bibfnamefont{Y.-S.} \bibnamefont{Liu}},
  \bibinfo{author}{\bibfnamefont{P.}~\bibnamefont{Yao}},
  \bibinfo{author}{\bibfnamefont{P.-A.} \bibnamefont{Glans}},
  \bibinfo{author}{\bibfnamefont{J.}~\bibnamefont{Guo}},
  \bibinfo{author}{\bibfnamefont{C.}~\bibnamefont{Chang}},
  \bibinfo{author}{\bibfnamefont{Z.}~\bibnamefont{Wu}},
  \bibinfo{author}{\bibfnamefont{W.}~\bibnamefont{Sakamoto}},
  \bibinfo{author}{\bibfnamefont{N.}~\bibnamefont{Itoh}},
  \bibinfo{author}{\bibfnamefont{T.}~\bibnamefont{Shimura}},
  \bibnamefont{et~al.}, \bibinfo{journal}{Phys. Rev. B}, \bibinfo{year}{2008},
  \textbf{\bibinfo{volume}{78}}, \bibinfo{pages}{085106}.

\bibitem[{\citenamefont{Henkelman et~al.}(2006)\citenamefont{Henkelman,
  Arnaldsson and J{\'o}nsson}}]{Bader1}
\bibinfo{author}{\bibfnamefont{G.}~\bibnamefont{Henkelman}},
  \bibinfo{author}{\bibfnamefont{A.}~\bibnamefont{Arnaldsson}}
  \bibnamefont{and}
  \bibinfo{author}{\bibfnamefont{H.}~\bibnamefont{J{\'o}nsson}},
  \bibinfo{journal}{Comput. Mater. Sci.}, \bibinfo{year}{2006}, 
  \textbf{\bibinfo{volume}{36}},
  \bibinfo{pages}{254}.

\bibitem[{\citenamefont{Sanville et~al.}(2007)\citenamefont{Sanville, Kenny,
  Smith and Henkelman}}]{Bader2}
\bibinfo{author}{\bibfnamefont{E.}~\bibnamefont{Sanville}},
  \bibinfo{author}{\bibfnamefont{S.~D.} \bibnamefont{Kenny}},
  \bibinfo{author}{\bibfnamefont{R.}~\bibnamefont{Smith}} \bibnamefont{and}
  \bibinfo{author}{\bibfnamefont{G.}~\bibnamefont{Henkelman}},
  \bibinfo{journal}{J. Comp. Chem.}, \bibinfo{year}{2007}, \textbf{\bibinfo{volume}{28}},
  \bibinfo{pages}{899}.

\bibitem[{\citenamefont{Tang et~al.}(2009)\citenamefont{Tang, Sanville and
  Henkelman}}]{Bader3}
\bibinfo{author}{\bibfnamefont{W.}~\bibnamefont{Tang}},
  \bibinfo{author}{\bibfnamefont{E.}~\bibnamefont{Sanville}} \bibnamefont{and}
  \bibinfo{author}{\bibfnamefont{G.}~\bibnamefont{Henkelman}},
  \bibinfo{journal}{J. Phys.: Condens. Matter}, \bibinfo{year}{2009},
   \textbf{\bibinfo{volume}{21}},
  \bibinfo{pages}{084204}.

\bibitem[{\citenamefont{Ravindran et~al.}(2006)\citenamefont{Ravindran, Vidya,
  Kjekshus, Fjellv{\aa}{a}g and Eriksson}}]{Eriksson}
\bibinfo{author}{\bibfnamefont{P.}~\bibnamefont{Ravindran}},
  \bibinfo{author}{\bibfnamefont{R.}~\bibnamefont{Vidya}},
  \bibinfo{author}{\bibfnamefont{A.}~\bibnamefont{Kjekshus}},
  \bibinfo{author}{\bibfnamefont{H.}~\bibnamefont{Fjellvag}}
  \bibnamefont{and} \bibinfo{author}{\bibfnamefont{O.}~\bibnamefont{Eriksson}},
  \bibinfo{journal}{Phys. Rev. B}, \bibinfo{year}{2006}, \textbf{\bibinfo{volume}{74}},
  \bibinfo{pages}{224412}.

\bibitem[{\citenamefont{Cor\`{a} et~al.}(2004)\citenamefont{Cor\`{a},
  Alfredsson, Mallia, Middlemiss, Mackrodt, Dovesi and Orlando}}]{cora1}
\bibinfo{author}{\bibfnamefont{F.}~\bibnamefont{Cor\`{a}}},
  \bibinfo{author}{\bibfnamefont{M.}~\bibnamefont{Alfredsson}},
  \bibinfo{author}{\bibfnamefont{G.}~\bibnamefont{Mallia}},
  \bibinfo{author}{\bibfnamefont{D.~S.} \bibnamefont{Middlemiss}},
  \bibinfo{author}{\bibfnamefont{W.~C.} \bibnamefont{Mackrodt}},
  \bibinfo{author}{\bibfnamefont{R.}~\bibnamefont{Dovesi}} \bibnamefont{and}
  \bibinfo{author}{\bibfnamefont{R.}~\bibnamefont{Orlando}},
  \bibinfo{journal}{Struct. Bond.}, \bibinfo{year}{2004}, \textbf{\bibinfo{volume}{113}},
  \bibinfo{pages}{171}.

\bibitem[{\citenamefont{Cor\`{a}}(2005)}]{cora2}
\bibinfo{author}{\bibfnamefont{F.}~\bibnamefont{Cor\`{a}}},
  \bibinfo{journal}{Mol. Phys.}, \bibinfo{year}{2005}, \textbf{\bibinfo{volume}{103}},
  \bibinfo{pages}{2483}.

\bibitem[{\citenamefont{Lebeugle et~al.}(2007)\citenamefont{Lebeugle, Colson,
  Forget and Viret}}]{BFOexp}
\bibinfo{author}{\bibfnamefont{D.}~\bibnamefont{Lebeugle}},
  \bibinfo{author}{\bibfnamefont{D.}~\bibnamefont{Colson}},
  \bibinfo{author}{\bibfnamefont{A.}~\bibnamefont{Forget}} \bibnamefont{and}
  \bibinfo{author}{\bibfnamefont{M.}~\bibnamefont{Viret}},
  \bibinfo{journal}{Appl. Phys. Lett.}, \bibinfo{year}{2007}, \textbf{\bibinfo{volume}{91}},
  \bibinfo{pages}{022907}.

\bibitem[{\citenamefont{Singh et~al.}(2005)\citenamefont{Singh, Ghita, Halilov
  and Fornari}}]{Singh1}
\bibinfo{author}{\bibfnamefont{D.~J.} \bibnamefont{Singh}},
  \bibinfo{author}{\bibfnamefont{M.}~\bibnamefont{Ghita}},
  \bibinfo{author}{\bibfnamefont{S.~V.} \bibnamefont{Halilov}}
  \bibnamefont{and} \bibinfo{author}{\bibfnamefont{M.}~\bibnamefont{Fornari}},
  \bibinfo{journal}{J. Phys. IV France}, \bibinfo{year}{2005}, \textbf{\bibinfo{volume}{128}},
  \bibinfo{pages}{47}.

\bibitem[{\citenamefont{Singh et~al.}(2006)\citenamefont{Singh, Ghita, Fornari
  and Halilov}}]{Singh2}
\bibinfo{author}{\bibfnamefont{D.~J.} \bibnamefont{Singh}},
  \bibinfo{author}{\bibfnamefont{M.}~\bibnamefont{Ghita}},
  \bibinfo{author}{\bibfnamefont{M.}~\bibnamefont{Fornari}} \bibnamefont{and}
  \bibinfo{author}{\bibfnamefont{S.~V.} \bibnamefont{Halilov}},
  \bibinfo{journal}{Ferroelectrics}, \bibinfo{year}{2006}, \textbf{\bibinfo{volume}{338}},
  \bibinfo{pages}{73}.

\bibitem[{\citenamefont{Ghita et~al.}(2005)\citenamefont{Ghita, Fornari, Singh
  and Halilov}}]{Singh3}
\bibinfo{author}{\bibfnamefont{M.}~\bibnamefont{Ghita}},
  \bibinfo{author}{\bibfnamefont{M.}~\bibnamefont{Fornari}},
  \bibinfo{author}{\bibfnamefont{D.~J.} \bibnamefont{Singh}} \bibnamefont{and}
  \bibinfo{author}{\bibfnamefont{S.~V.} \bibnamefont{Halilov}},
  \bibinfo{journal}{Phys. Rev. B}, \bibinfo{year}{2005}, \textbf{\bibinfo{volume}{72}},
  \bibinfo{pages}{054114}.

\bibitem[{\citenamefont{Suewattana and Singh}(2006)}]{Singh4}
\bibinfo{author}{\bibfnamefont{M.}~\bibnamefont{Suewattana}} \bibnamefont{and}
  \bibinfo{author}{\bibfnamefont{D.~J.} \bibnamefont{Singh}},
  \bibinfo{journal}{Phys. Rev. B}, \bibinfo{year}{2006}, \textbf{\bibinfo{volume}{73}},
  \bibinfo{pages}{224105}.

\bibitem[{\citenamefont{Yamauchi et~al.}(2008)\citenamefont{Yamauchi, Freimuth,
  Bluegel and Picozzi}}]{HMO4}
\bibinfo{author}{\bibfnamefont{K.}~\bibnamefont{Yamauchi}},
  \bibinfo{author}{\bibfnamefont{F.}~\bibnamefont{Freimuth}},
  \bibinfo{author}{\bibfnamefont{S.}~\bibnamefont{Bluegel}} \bibnamefont{and}
  \bibinfo{author}{\bibfnamefont{S.}~\bibnamefont{Picozzi}},
  \bibinfo{journal}{Phys. Rev. B}, \bibinfo{year}{2008}, \textbf{\bibinfo{volume}{78}},
  \bibinfo{pages}{014403}.

\bibitem[{\citenamefont{Alonso et~al.}(2000)\citenamefont{Alonso,
  Mart{\'i}nez-Lopez, Casais and Fern{\'a}ndez-D{\'i}az}}]{HMO3}
\bibinfo{author}{\bibfnamefont{J.~A.} \bibnamefont{Alonso}},
  \bibinfo{author}{\bibfnamefont{M.~J.} \bibnamefont{Mart{\'i}nez-Lopez}},
  \bibinfo{author}{\bibfnamefont{M.~T.} \bibnamefont{Casais}}
  \bibnamefont{and} \bibinfo{author}{\bibfnamefont{M.~T.}
  \bibnamefont{Fern{\'a}ndez-D{\'i}az}}, \bibinfo{journal}{Inorg. Chem.}, \bibinfo{year}{2000},
  \textbf{\bibinfo{volume}{39}}, \bibinfo{pages}{917}.

\bibitem[{\citenamefont{Lufaso and Woodward}(2004)}]{Tilting}
\bibinfo{author}{\bibfnamefont{M.~W.} \bibnamefont{Lufaso}} \bibnamefont{and}
  \bibinfo{author}{\bibfnamefont{P.~M.} \bibnamefont{Woodward}},
  \bibinfo{journal}{Acta Cryst. B}, \bibinfo{year}{2004}, \textbf{\bibinfo{volume}{60}},
  \bibinfo{pages}{10}.

\bibitem[{\citenamefont{Tachibana et~al.}(2007)\citenamefont{Tachibana,
  Shimoyama, Kawaji, Atake and T.-Muromachi}}]{Tilt2}
\bibinfo{author}{\bibfnamefont{M.}~\bibnamefont{Tachibana}},
  \bibinfo{author}{\bibfnamefont{T.}~\bibnamefont{Shimoyama}},
  \bibinfo{author}{\bibfnamefont{H.}~\bibnamefont{Kawaji}},
  \bibinfo{author}{\bibfnamefont{T.}~\bibnamefont{Atake}} \bibnamefont{and}
  \bibinfo{author}{\bibfnamefont{E.}~\bibnamefont{Takayama-Muromachi}},
  \bibinfo{journal}{Phys. Rev. B}, \bibinfo{year}{2007}, \textbf{\bibinfo{volume}{75}},
  \bibinfo{pages}{144425}.

\bibitem[{\citenamefont{Martin}(2004)}]{MartinBook}
\bibinfo{author}{\bibfnamefont{R.~L.} \bibnamefont{Martin}},
  \emph{\bibinfo{title}{Electronic Structure: Basic Theory and Methods}}
  (\bibinfo{publisher}{Cambridge University Press}, \bibinfo{year}{2004}).

\bibitem[{\citenamefont{Sergienko et~al.}(2006)\citenamefont{Sergienko,
  C.\c{S}en and Dagotto}}]{dagotto}
\bibinfo{author}{\bibfnamefont{I.~A.} \bibnamefont{Sergienko}},
  \bibinfo{author}{\bibnamefont{C.\c{S}en}} \bibnamefont{and}
  \bibinfo{author}{\bibfnamefont{E.}~\bibnamefont{Dagotto}},
  \bibinfo{journal}{Phys. Rev. Lett.}, \bibinfo{year}{2006}, \textbf{\bibinfo{volume}{97}},
  \bibinfo{pages}{227204}.

\bibitem[{\citenamefont{Lorenz et~al.}(2007)\citenamefont{Lorenz, Wang and
  Chu}}]{Lorenz}
\bibinfo{author}{\bibfnamefont{B.}~\bibnamefont{Lorenz}},
  \bibinfo{author}{\bibfnamefont{Y.-Q.} \bibnamefont{Wang}} \bibnamefont{and}
  \bibinfo{author}{\bibfnamefont{C.~W.} \bibnamefont{Chu}},
  \bibinfo{journal}{Phys. Rev. B}, \bibinfo{year}{2007}, \textbf{\bibinfo{volume}{76}},
  \bibinfo{pages}{104405}.

\bibitem[{\citenamefont{Pomjakushin et~al.}(2009)\citenamefont{Pomjakushin,
  Kenzelmann, Donni, Harris, Nakajima, Mitsuda, Tachibana, Keller, Mesot,
  Kitazawa et~al.}}]{Yu}
\bibinfo{author}{\bibfnamefont{V.~Y.} \bibnamefont{Pomjakushin}},
  \bibinfo{author}{\bibfnamefont{M.}~\bibnamefont{Kenzelmann}},
  \bibinfo{author}{\bibfnamefont{A.}~\bibnamefont{Donni}},
  \bibinfo{author}{\bibfnamefont{A.~B.} \bibnamefont{Harris}},
  \bibinfo{author}{\bibfnamefont{T.}~\bibnamefont{Nakajima}},
  \bibinfo{author}{\bibfnamefont{S.}~\bibnamefont{Mitsuda}},
  \bibinfo{author}{\bibfnamefont{M.}~\bibnamefont{Tachibana}},
  \bibinfo{author}{\bibfnamefont{L.}~\bibnamefont{Keller}},
  \bibinfo{author}{\bibfnamefont{J.}~\bibnamefont{Mesot}},
  \bibinfo{author}{\bibfnamefont{H.}~\bibnamefont{Kitazawa}},
  \bibnamefont{et~al.}, \bibinfo{journal}{New J. Phys.}, \bibinfo{year}{2009},
  \textbf{\bibinfo{volume}{11}}, \bibinfo{pages}{043019}.

\bibitem[{\citenamefont{Ederer and Spaldin}(2005)}]{BFO.LDAU}
\bibinfo{author}{\bibfnamefont{C.}~\bibnamefont{Ederer}} \bibnamefont{and}
  \bibinfo{author}{\bibfnamefont{N.~A.} \bibnamefont{Spaldin}},
  \bibinfo{journal}{Phys. Rev. B}, \bibinfo{year}{2005}, \textbf{\bibinfo{volume}{71}},
  \bibinfo{pages}{224103}.

\bibitem[{\citenamefont{Koichi and Fujio}(2008)}]{vesta}
\bibinfo{author}{\bibfnamefont{M.}~\bibnamefont{Koichi}} \bibnamefont{and}
  \bibinfo{author}{\bibfnamefont{I.}~\bibnamefont{Fujio}}, \bibinfo{journal}{J.
  Appl. Cryst.}, \bibinfo{year}{2008}, \textbf{\bibinfo{volume}{41}}, \bibinfo{pages}{653}.

\bibitem[{\citenamefont{Perdew et~al.}(2008)\citenamefont{Perdew, Ruzsinszky,
  Csonka, Vydrov, Scuseria, Constantin, Zhou and Burke}}]{PBEsol}
\bibinfo{author}{\bibfnamefont{J.~P.} \bibnamefont{Perdew}},
  \bibinfo{author}{\bibfnamefont{A.}~\bibnamefont{Ruzsinszky}},
  \bibinfo{author}{\bibfnamefont{G.~I.} \bibnamefont{Csonka}},
  \bibinfo{author}{\bibfnamefont{O.~A.} \bibnamefont{Vydrov}},
  \bibinfo{author}{\bibfnamefont{G.~E.} \bibnamefont{Scuseria}},
  \bibinfo{author}{\bibfnamefont{L.~A.} \bibnamefont{Constantin}},
  \bibinfo{author}{\bibfnamefont{X.}~\bibnamefont{Zhou}} \bibnamefont{and}
  \bibinfo{author}{\bibfnamefont{K.}~\bibnamefont{Burke}},
  \bibinfo{journal}{Phys. Rev. Lett.}, \bibinfo{year}{2008}, \textbf{\bibinfo{volume}{100}},
  \bibinfo{pages}{136406}.

\bibitem[{\citenamefont{Sosnowska et~al.}(2002)\citenamefont{Sosnowska,
  Schafer, Kockelmann Andersen and Troyanchuk}}]{exp1}
\bibinfo{author}{\bibfnamefont{I.}~\bibnamefont{Sosnowska}},
  \bibinfo{author}{\bibfnamefont{W.}~\bibnamefont{Schafer}},
  \bibinfo{author}{\bibfnamefont{W.}~\bibnamefont{Kockelmann}},
  \bibinfo{author}{\bibfnamefont{K.~H.} \bibnamefont{Andersen}}
  \bibnamefont{and} \bibinfo{author}{\bibfnamefont{I.~O.}
  \bibnamefont{Troyanchuk}}, \bibinfo{journal}{Appl. Phys. A: Mater. Sci.
  Process.}, \bibinfo{year}{2002}, \textbf{\bibinfo{volume}{74}}, \bibinfo{pages}{S1040}.

\end{thebibliography}

\newpage

\newpage
\begin{table}[htbp]
   \centering
   \begin{tabular}{lcccc}
 \hline
                    &  PBE       & B1-WC    & HSE      & Exp      \\
\hline
a$_ {rh}$ (\AA)     &5.687      &  5.609   & 5.651    & 5.634  \\
$\alpha_{rh}$ (deg) &59.22      &  59.37   & 59.12    & 59.35  \\
V (\AA$^{3}$)       &127.79     &  122.99  & 125.04   & 124.60 \\
\hline
$x_{Fe}$            &0.223      & 0.219    & 0.219     &  0.221   \\
$x_{O} $        &0.533      & 0.511    & 0.522     &  0.538   \\
$y_{O} $            &0.936      & 0.926    & 0.931     &  0.933   \\
$z_{O}$             &0.387      & 0.406    & 0.394     &  0.395   \\
\hline
d$^{l}_{Bi-O}$  (\AA)&2.492     & 2.591    & 2.533     &  2.509   \\
d$^{s}_{Bi-O}$ (\AA) &2.308     & 2.196    & 2.249     &  2.270   \\
d$^{l}_{Fe-O}$ (\AA) &2.164     & 2.102    & 2.121     &  2.110   \\
d$^{s}_{Fe-O}$ (\AA) &1.957     & 1.932    & 1.948     &  1.957   \\
$\beta_{O-Fe-O}$ (deg) &164.09  &162.84    & 164.56    & 165.03   \\
 \hline
$\mu_{Fe}$($\mu_{B}$) &   3.72  & 4.2      & 4.12      &   3.75    \\
Gap (eV)              &   1.0   &3.0        & 3.4      & 2.5-2.8  \\
\hline

\end{tabular}
\caption{Dependence of structural paramenters, magnetic moment and electronic band gap with respect to PBE (this work), B1-WC (Ref.\cite{Goffinet}) and HSE (this work). a$_ {rh}$, $\alpha_{rh}$,
V  denote  the lattice parameter, the angle, and the volume of BiFeO$_{3}$ in space group $R3c$,  in the rhombohedral setting. Fe ($2a$) and O ($6b$)  Wyckoff positions are also reported. Bi is in the  ($2a$) (0,0,0) position (not reported in the table). Long (l) and short (s) bond lengths of Bi-O and Fe-O are given. $\beta_{O-Fe-O}$ is the  O-Fe-O obtuse angle  (equal to 180$^{\circ}$ in the ideal octahedron). $\mu_{Fe}$ is Fe magnetic moment. Experimental structural parameters are from Ref.\cite{Kubel}, the experimental magnetic moment (measured on polycrystalline powder) is from Ref.\ \cite{exp1}; the band gap is from Refs.\cite{gap1,gap2,gap3,gap4,Kanai,Gao}.}
\label{tab1}
\end{table}

\newpage
\begin{table}[htbp]
   \centering
   \begin{tabular}{lcc}
 \hline

\multicolumn{3}{c}{BiFeO$_{3}$ ($R3c$)}\\
BiFeO$_{3}$ &  Q$_{B}$($e$)        &  V$_{B}$ (\AA$^{3}$)   \\
\hline
Bi          &  13.13  (13.01)      &    18.06 (17.67)  \\
Fe          &  12.38  (12.13)      &    8.24 (7.74)   \\
O           &   7.16  (7.28)       &    12.08 (12.38)  \\
\hline
\multicolumn{3}{c}{HoMnO$_{3}$ (AFM-E)}\\
Ho          &    5.18  (5.00)      &  7.41  (6.77)    \\
Mn          &    6.82  (6.75)      &  12.10 (11.84)  \\
O           &    7.33  (7.41)      &  12.42 (12.61) \\
 \hline

\end{tabular}
\caption{Calculated charge and volumes
according to Bader AIM partitioning
for a fixed HSE geometry for BiFeO$_{3}$ and HoMnO$_{3}$.
Numbers in parenthesis are based on the HSE charge density.
In all cases, we used the ferroelectric phase. }
\label{bader}
\end{table}

\begin{table}
 \begin{center}
\begin{tabular}{lccc}

\hline
 AFM-A             &PBE      & HSE    & Exp.  \\
Ho $4c(x,\frac{1}{4},z)$ &&&\\
$x$              &0.0856  &0.0859 & 0.0839       \\
$z$              &0.4805  &0.4816 & 0.4825       \\
Mn $4a (0 0 0 )$ &  &            &                \\
O$_{1}$ $4c(x \frac{1}{4}z)$&&&\\
 $x$              &0.4617 & 0.4616&  0.4622     \\
 $z$              &0.6162 & 0.6179&  0.6113     \\
O$_{2}$ $8d(x y z)$     &&&\\
$x$               &0.3250 &0.3301& 0.3281 \\
$y$               &0.0550 &0.0573& 0.0534  \\
$z$               &0.1988 &0.2018& 0.2013 \\
\hline
Mn-O  (s-inpl)  & 1.9271  &  1.9022  & 1.9044   \\
Mn-O  (l-inpl)  & 2.2030 &  2.2393  &  2.2226   \\
Mn-O (outpl)   &  1.9518  & 1.9546   &   1.9435  \\
Q                &0.55    & 0.61     & 0.60 \\
\hline
$\alpha$         & 19.47   &  19.71       &   18.77   \\
Mn-O-Mn (inpl)   & 143.849   &  142.839    &  144.081  \\
Mn-O-Mn (outpl)  &  141.060   &  140.586    &  142.462  \\
\hline
\end{tabular}
\caption{Structural parameters for HoMnO$_{3}$ optimized with AFM-A
configuration in the $Pnma$ unit cell. The lattice parameters
are $a$=5.8354,$b$=7.3606,$c$=5.2572 \AA.
Long and small Mn-O distances in $c-a$ plane (inpl) and
middle Mn-O distance along the $b$ axis (outpl).
Mn-O-Mn angles in the $c-a$ plane  (inpl) and interplane
Mn-O-Mn angles (with apical O along the $b$ axis).}
\label{tab.afma}
\end{center}
 \end{table}

%

\begin{table}[htbp]
\centering
\begin{tabular}{lcccc}
\hline
BiFeO$_{3}$& P$_{ionic}$    & P$_{ele}$  & P$_{tot}$  & P$_{pcm}$ \\

PBE        &   171.1       & $-$65.5     &  105.6  &   87.8   \\
HSE        &   177.4       & $-$67.1     &  110.3  &  103.4   \\
PBE(HSE)   &   177.4      & $-$64.8      &  112.6  &  103.4   \\
HSE(PBE)   &   171.1       & $-$67.9     &  103.2  &   87.8   \\
\hline
HoMnO$_{3}$&     &  &    &  \\
\hline
PBE        & $-$0.6       &  $-$5.2   &   $-$5.8  &  $-$2.0  \\
HSE        & $-$0.3       &  $-$1.6   &   $-$1.9  &  $-$0.9  \\
PBE(HSE)   & $-$0.3       &  $-$3.1   &   $-$3.4  &  $-$0.9 \\
HSE(PBE)   & $-$0.6       &  $-$2.9   &   $-$3.5  &  $-$2.0  \\
\hline
\end{tabular}

\caption{Ferroelectric polarization  of BiFeO$_{3}$ and HoMnO$_{3}$,
using PBE and  HSE and Point Charge Model estimate is also given (P$_{pcm}$). 
Units are in $\mu C /cm^{2}$.
PBE (HSE) means a PBE calculation at fixed HSE geometry
for the ferroelectric as well as for the paraelectric structure.
Viceversa for HSE (PBE). For BiFeO$_{3}$ the polarization is along the [111] direction; for HoMnO$_{3}$ is along the $c$ axis. The partial ionic and electronic contributions are also given. 
 Obviously, the relative weigth of ionic and electronic contributions  depends on the
valence electronic configuration, \textit{e.g.} Bi $d$ in the core or in the valence. Therefore, given an electronic configuration,
we are only interested in trends of P$_{tot}$ and P$_{pcm}$ 
as far as the exchange-correlation functional is concerned.}
\label{tab4}
\end{table}
\newpage

\newpage

\newpage

\newpage

\newpage

\newpage

\end{document}